\documentclass[aps,prb,twocolumn,floatfix]{revtex4-1}
\usepackage{graphics}
\usepackage{subfigure}
\usepackage{epsfig}
\usepackage{epsf,epic}
\usepackage{color}
\usepackage{marvosym }
\usepackage{subfigure}
\usepackage{amsmath}
\usepackage{amssymb}
\usepackage{amsfonts}
\usepackage{wrapfig}
\usepackage{multirow}
\usepackage{bm}
\usepackage{dcolumn}% Align table columns on decimal point
\newcommand{\etal}{\textit{et al.\ }}
\newcommand{\ie}{\textit{i.e.\ }}
\newcommand{\eg}{\textit{e.g.\ }}
\begin{document}
\title{Computational identification of Ga-vacancy related electron paramagnetic resonance  centers in $\beta$-Ga$_2$O$_3$}
\author{Dmitry Skachkov and Walter R. L. Lambrecht}
\affiliation{Department of Physics, Case Western Reserve University, 10900 Euclid Avenue, Cleveland, OH-44106-7079, U.S.A.}
\author{Hans J\"urgen von Bardeleben}
\affiliation{Sorbonne Universit\'es, UPMC Universit\'e Paris 06, UMR7588, Institut des Nanosciences de Paris, 4 place Jussieu, 75005 Paris, France}
\author{Uwe Gerstmann}
\affiliation{Lehrstuhl f\"ur Theoretische Materialphysik, Universit\"at Paderborn, 33098 Paderborn, Germany}
\author{Quoc Duy Ho and Peter De\'ak}
\affiliation{Bremen Center for Computational Materials Science, University of Bremen, P.O. Box 330440, D-28334 Bremen, Germany}

\begin{abstract}
  A combined experimental/theoretical study of the electron paramagnetic centers
  in irradiated $\beta$-Ga$_2$O$_3$ is presented.  Four EPR spectra,
  two $S=1/2$ and two $S=1$, are observed after high-energy proton or electron
  irradiation. Three of them have been  reported before in neutron irradiated
  samples. One of the S=1/2 spectra (EPR1) can be observed at room temperature
  and below and  is characterized by the spin Hamiltonian parameters
  $g_b=2.0313$, $g_c=2.0079$, $g_{a*}=2.0025$ and a quasi isotropic hyperfine interaction
  with two equivalent Ga neighbors of $~\sim$14 G on $^{69}$Ga and correspondingly $\sim$18 G on $^{71}$Ga in their natural
  abundances. The second (EPR2) is observed after photoexcitation
  (with threshold 2.8 eV) at low temperature and is characterized
  by $g_b=2.0064$, $g_c=2.0464$, $g_{a*}=2.0024$ and
  a quasi isotropic hyperfine interaction with two equivalent Ga neighbors of 10 G (for $^{69}$Ga).
  A spin $S = 1$ spectrum with a similar g-tensor and a 50\% reduced  hyperfine splitting accompanies each of these,
  which is indicative of a defect of two weakly coupled $S=1/2$ centers.
  Density functional theory calculations of the magnetic resonance
  fingerprint ($g$-tensor and hyperfine interaction)
  of a wide variety  of native defect models and their
complexes are carried out to identify these EPR centers in terms of
specific defect configurations. The EPR1 center is proposed to
correspond to a complex of two tetrahedral $V_\mathrm{Ga1}$ with an
interstitial Ga in between them and oriented in a specific direction
in the crystal. This model was previously shown to have lower energy
than the simple tetrahedral Ga vacancy and has a $2-/3-$ transition level higher than
other $V_\mathrm{Ga}$ related models, which would explain
why the other ones are already in their diamagnetic $3-$ state  and are thus
not observed if the Fermi level is pinned
approximately at this level. The EPR2 spectra ($S=1/2$ as well as the related $S=1$) are proposed
to correspond to the octahedral
$V_\mathrm{Ga2}$ in which the spin is located on an oxygen off the defect's
mirror plane  and has a tilted spin density.
Models based on self-trapped holes and oxygen interstitials
are ruled out because they would have hyperfine interaction with more than
two Ga nuclei  and because they can not support a corresponding $S=1$ center.
\end{abstract}
\maketitle
\section{Introduction}\label{sec:intro}
Monoclinic $\beta$-Ga$_2$O$_3$ has recently
attracted attention as an ultra-wide-band-gap
semiconductor.\cite{Sasaki13}
Its band gap of about 4.7 eV \cite{Matsumoto74,Peelaers15a,Furthmuller16, Mengle16,Ratnaparkhe17} combined with unintentionally doped
semiconducting rather than insulating properties make it attractive
for high-power electronics applications. Mostly, the wide band gap leads to a
high breakdown field (estimated to be possibly as high as 8 MV/cm based
on the relation between band gap and break down voltage in other materials,
and already demonstrated\cite{Green16} to be as high as 3.8 MV/cm), plays an important role in various figures of merit (FOM) for high-power transistor design,
such as Baliga's FOM.\cite{Baliga}
Its good transparency in the ultraviolet region also make it
suitable as a transparent conductor.\cite{Peelaers15}
The origin of unintentional doping and the limitations on the degree of n-type
doping that can be achieved depends on a thorough understanding of the
defect physics. While a substantial amount of work
\cite{Varley11,Varley12,Harwig78,Zacherle13,Peelaers16,Deak17,Ingebrigtsen,Irmscher11,Zhang16,Ingebrigtsen19,Gao18,Islam19,Kananen17,Kananensth,Jurgen18} has already appeared on the defect
physics, the experimental
signatures of many of the defects are still unclear.

\begin{figure}
  \includegraphics[width=8cm]{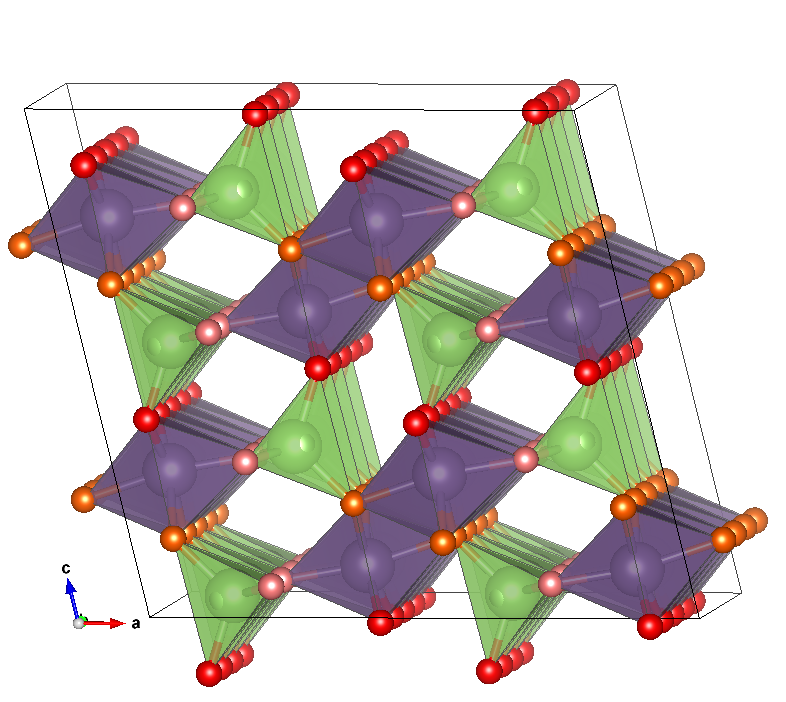}
  \caption{Crystal structure of $\beta$-Ga$_2$O$_3$  shown in
    the 160 atom supercell indicating the
    polyhedra surrounding tetrahedral Ga$_{(1)}$ (in green) and
    octahedral Ga$_{(2)}$ (purple). The O$_{(1)}$, O$_{(2)}$, O$_{(3)}$
    are colored coded as red, pink and orange.\label{figstruc}}
\end{figure}

Recently, an Electron Paramagnetic Resonance (EPR) 
center was reported by Kananen \etal\cite{Kananen17} in neutron irradiated
samples and ascribed to the octahedral Ga-vacancy site. 
In $\beta$-Ga$_2$O$_3$ there are two nonequivalent Ga sites one
with a tetrahedral coordination (Ga$_{(1)}$) and one with an octahedral
coordination (Ga$_{(2)}$).\cite{Geller60}  Likewise, there are three distinct
O sites, O$_{(1)}$ and O$_{(2)}$ are each connected to three Ga while O$_{(3)}$
is connected to four Ga (see Fig. \ref{figstruc}).

Here we present an EPR study of  $\beta$-Ga$_2$O$_3$ with defects 
introduced by high energy particle irradiation combined with
first-principles calculations of the hyperfine interaction (HFI) and
gyromagnetic $g$-tensors.
A similar spectrum to that of Ref. \onlinecite{Kananen17} is observed
after irradiation in the dark. 
After photoexcitation a different center appears with
different orientation of the main $g$-tensor axes and slightly smaller
HFI-values. That EPR center has properties similar to those previously ascribed
to the self-trapped hole (STH),\cite{Kananensth} which similarly
has spin density localized on an oxygen $p$-orbital but would not involve
a Ga-vacancy. It was previously observed after X-ray irradiation
at low temperature on already neutron irradiated samples.
Two distinct $S=1$ spectra are also reported here.  The first of these,
related to EPR1, was
previously reported,\cite{Kananen17} the second, related to EPR2,  was not.
The characteristic 
$g$-tensors and hyperfine splitting of these centra and their thermal stability 
are  presented. 

Second, we present first-principles calculations for a wide variety of native defect
models  with the goal of identifying the chemical nature
of the observed defect centers. 
The paper is organized as follows. We first present 
our computational methodology in Sec. \ref{sec:method} and give details of the experiment
in Sec. \ref{sec:exp}.
Next, we present the experimental results in Sec. \ref{sec:exptresults}. Additional information
on the experiments can be found in a brief prior report of this work.\cite{Jurgen18}
The models investigated and the reasons why are described in Sec.  \ref{sec:models},
followed by a presentation of the computational results in Sec. \ref{sec:compresults}.
We present a thorough comparison of the experimental results with
calculations in Sec. \ref{sec:discussion} and from it deduce the most likely models. 
A summary completes the paper.

\section{Computational Methods}\label{sec:method}
Density functional theory (DFT) is used in this work  to first relax the structure of
the models and determine their transition levels. Subsequently the EPR parameters are calculated using the
Gauge Including Projector Augmented Wave (GIPAW) method.

For a proper description of the strongly localized electron in the oxygen
dangling bonds, structural relaxation of the models
has been done with the  hybrid functional \cite{PBEh,HSE03,HSE06} with the parameters chosen
as in Ref. \onlinecite{Deak17}. Specifically,  the fraction of non-local exchange
was $\alpha=0.26$ with screening parameter $\mu=0.00$.\cite{Deak17} These parameters
were determined in that paper to optimize both the band gap and satisfaction of the
Generalized Koopmans Theorem (GKT) for various defects. They are in fact
very close to the original PBE0\cite{PBEh} parameters. This portion of the work used 
the Vienna Ab-initio Simulation Package (VASP).\cite{Kressevasp1,Kressevasp3,vasp}
The same methodology as in Ref. \onlinecite{Deak17} was also used to find the transition levels of the
defects including details of the image charge corrections for charged states of the defect.
In fact, for the models already studied in Ref. \onlinecite{Deak17} we
use the results directly from that paper while for the additional models the same methodology
was followed here.

The GIPAW method\cite{Pickard01,Pickard02} uses self-consistent density
functional perturbation theory to calculate the linear magnetic response
of the defect system to an  external magnetic field and is used
here in its implementation
included in the Quantum Espresso package.\cite{QE-2009}
The determination of the $g$-tensor via the GIPAW method is presently restricted
to (semi-)local functionals: The non-local exact-exchange terms specific to hybrid functional 
are not yet included in the $g$-tensor code. Thus, we used
a combined approach, where the structures previous relaxed within hybrid functional are then kept
fixed and their electronic structure is recalculated at the GGA level
using the Perdew-Burke-Ernzerhof (PBE)\cite{PBE} functional and the wave functions and spin-densities
obtained in this manner are then used to evaluate the $g$-tensor using the GIPAW code.
We found that once the structure is relaxed with hybrid functional, the spin density
recalculated with GGA-PBE without the hybrid functional terms is quite similar to the hybrid
functional one and stays well localized on a single oxygen in all models we investigated.
Thus, this approach should be adequate to determine the $g$-tensor.

A few comments are in order here on the accuracy with which we expect to be able to predict $g$-tensors. 
Agreement between calculated and experimental data for the principal
 values of the $g$-tensor is obtained typically to better than 0.0003 at the semilocal DFT
 level.\cite{Bardeleben12prl,Bardeleben14,Pfanner12,Pickard02,George13,Rohrmuller17}
 However these errors are on a small $g$-tensor shift from the free-electron value and
 amount to a relative error of 30 \%.  In some cases, such as the Jahn-Teller distorted split interstitial
 (N-N$_\mathrm{N}^0$) in GaN the deviation for  a single specific directions was 0.007.\cite{Bardeleben14}
 All these defects have well localized spins already at the GGA level.  There is not
 much experience yet with $g$-tensor calculations for acceptor type defects
 for which the localizaton of the spin is sensitive to the functional because of self-interaction errors,
 as is the case here.  Thus, we estimate a conservative error bar of  0.01
 on the larger $\Delta g$-tensor components.   Therefore, 
 to achieve  our main goal of identifying the defect centers responsible for the observed EPR spectra,
 we prefer to focus on the qualitative aspects of the $g$-tensor,
 such as the ordering of the principal value directions. 

As far as the hyperfine interaction both the VASP and GIPAW program were used. This part of the
GIPAW program does not require the phase factors related to gauge-inclusion. The hyperfine tensor $A$ 
consists of the isotropic Fermi-contact term, which essentially requires calculating the spin density at the
 relevant nuclear sites which carry a nuclear spin, which involves the $s$-like part of the
wavefuncions on these sites only. In the present case, this corresponds to the Ga sites only because
O has an isotope without nuclear spin with more than 99.9\%  abundance.  We are thus dealing with
superhyperfine (SHF) interaction. As expected, the SHF intereaction is
predominantly isotropic both in the computations and experiment. Therefore, we focus on the Fermi contact term  and 
on the number of Ga atoms which exhibit a strong hyperfine interaction. 

Within the VASP code we use the hybrid functional  calculated spin density.
It was found that the hybrid  functional overestimates  the  experimental values  by about a
factor 2 for the $V_\mathrm{Ga}$ based models. The Fermi contact term clearly is very sensitive
to the degree of localization of the defect wave function. Compared to the PBE-GGA result, we found
that the wave function became a bit more localized in hybrid functional but it becomes more localized
on both the O and the Ga and thereby in fact slightly increases the HFI.  

Although the hybrid functional was specifically designed to satisfy the GKT and is thus
already expected to describe the {\em overall} localization accurately, we should note that
in calculations of the hyperfine intereaction, we are placing a higher demand on the
accuracy of the wave functions at the individual nuclei. The GKT requires linear dependence of the total energies
as function of occupation number of the defect state and focuses
on energy but for hyperfine it is the wave function itself that matters.
It must have the right shape and distribution over O and its Ga neighbors
and the right amount of Ga-$s$ character. We therefore also considered the DFT$+U$ approach
and studied how different values of $U$ affect the HFI. 
DFT$+U$ is often used to mimic the effects of non-local exchange of hybrid functionals. In
 the present case, the defect wave functions are primarily O-$p$ orbital derived
and hence applying on-site Coulomb or Hubbard $U$ terms on the O-$p$ orbitals may be expected
to make the spin-density related to the O-$p$ hole in the EPR active state more localized on O-$p$.
We should point out here that we apply such $U$ terms on all O but it affects mainly the
localization of the particular O on which the hole is localized because the empty states
in DFT$+U$ are affected differently from the filled ones. We find indeed that
using this DFT$+U$ approach with $U$ on O-$p$ provides relaxed structures similar to hybrid functional.
This is because the localization of the spin on a single atom is accompanied by related structural
distortion via a feedback loop. Thus the DFT$+U$ method can be used as  an alternative to
hybrid functional to obtain realistic structural models for these defects. 

However, {\sl a-priori} it is not clear whether this DFT$+U$ approach would also increase 
the spin-density on the neighboring Ga or reduce it. We find, in fact, that
it pulls weight away from the Ga toward the O and hence reduces the HFI on the Ga atoms. 
We estimated the size of $U$ using the linear response approach proposed by Cococcioni and de Gironcoli\cite{Cococcioni5,Timrov18} for some of the simple $V_\mathrm{Ga}$ 
defects and found values of about $U\approx6-7$ eV. 
For another defect with spin on a neighboring oxygen, Mg$_\mathrm{Ga}$, the HFI calculated within 
hybrid functional\cite{Ho18} or pure PBE agree well with experiment without adding $U$ terms. 
Both a Mg$_\mathrm{Ga}$ and a $V_\mathrm{Ga}$,
present a repulsive potential which pushes out a state above the valence band and which is
thus O-$p$-like.  However, one still expects it to have more of a dangling bond character
for $V_\mathrm{Ga}$ than for Mg$_\mathrm{Ga}$. Interstitial O$_i$ also correspond to
O-localized spins and are considered here and found to have somewhat smaller HFI on their
neighboring Ga than $V_\mathrm{Ga}$ within the same functional. We thus conclude that
a single $U$ value or hybrid functional may not be optimal for all defects.  
In the results below, we will start from a somewhat conservative value of $U=4$ eV
used for all models to allow a consistent comparison and
discuss how the results change with $U$ although we do not propose to
simply adjust $U$ until agreement is obtained.  Instead we base our conclusions on
identifying models mostly on qualitative aspects, namely how many Ga atoms have significant
hyperfine interaction. 

We have also considered another possible reason for the overestimate of the HFI.
For both EPR1 and EPR2 the most promising models turn out to undergo a symmetry breaking.
In the EPR1 case, it corresponds to a complex consisting of two vacancies and an interstitial
in between them which has an inversion center. For EPR2, the defect has a mirror plane.
The spin could thus be on either side of the defect but is found to be localized on
one side.  However, we may now consider that this symmetry breaking could undergo a dynamic 
Jahn-Teller effect. In other words, there is a coupling to a local vibrational mode which
presents a double well potential energy land scape. When the atoms are displaced one way, the
spin localizes on one side and vice versa. This is because of the feedback loop between localization
and atomic displacements mentioned earlier. The spin would then flip back and forth
between the two equivalent sides of the defect.  The total wave function is then a product
of the electronic wave function and the vibronic one and in calculating the probability
of the electron to be on a given nucleus, which gives the hyperfine interaction, one would
have to carry out an integral over the vibronic wave function modulo squared and take a thermal
average over the vibronic states. Unfortunately, it is beyond the scope of the present work to
evaluate this effect quantitatively. 
It nonetheless becomes clear that spin-phonon coupling could lead to a dynamical 
reduction factor of the HFI. In a classical picture, the spin spends part of its time
on one side and part of its time on the other side but also in transit between
the two and hence in a less localized state  so that ultimately the probability to find the
spin at the nuclear sites is reduced.  Such reduction factors due to dynamical Jahn-Teller
effects have been discussed in literature before, for example in the paper by Ham\cite{Ham65}
and in Mauger \etal\cite{Mauger87}. It would lead to a temperature dependent HFI, which
has not yet been studied in detail  but also could lead to an overall reduction.

In summary, although we will show below that the calculations overestimate the contact HFI,
we do not view this  as a crucial point in making our identification of the models. It
is tentatively attributed to the difficulty in describing the wave function of dangling
bond like acceptor states subject to strong self-interaction errors. Even though hybrid
functionals and DFT$+U$ methods can remedy this problem partially, it is not yet clear that a single
choice will describe different defects equally well. In addition, dynamical
Jahn-Teller effects, not yet included in the first-principles framework may reduce these
hyperfine factors and also affect spin-orbit coupling and $g$-tensors precise values.

Further details of the computational method are as follows. 
The defects were simulated using periodic boundary conditions in 
160 and 240  atom supercells, which are respectively
a $1\times4\times2$ and $1\times4\times3$ superlattices of the 20 atom
conventional cell of the base-centered $C2/m$-spacegroup
$\beta$-Ga$_2$O$_3$ structure.\cite{Geller60}
The plane wave expansion was used with a cut-off of 100 Ry and the
Brillouin zone integration used the $\Gamma$-point only for the
self-consistent calculations and hyperfine structure. Convergence
was tested by also using a $2\times2\times2$ shifted mesh. For the
$g$-tensor calculations, which are more sensitive to {\bf k}-point convergence, the $2\times2\times2$ mesh was used and convergence was tested by also using a
$3\times3\times3$ mesh in a few test cases.
Troullier-Martins type pseudopotentials obtained within the
PBE exchange correlation functional were used.

The hyperfine parameters were calculated using the QE-GIPAW code\cite{gipaw} as well as the VASP code. 
The relativistic hyperfine tensor consists of the isotropic Fermi contact
term which requires the spin density within a distance of the
Thomas radius ($r_T=Ze^2/mc^2$) from the nuclear sites as well as the dipolar terms.\cite{Blugel87} (Here, $Z$ is the atomic number, $e$ the elementary charge, $m$ the free electron mass and $c$ the speed of light.) Within a pseudopotential or projector augmented wave approach, a reconstruction of the all-electron wave functions from
the pseudo wave functions is required.\cite{VdWalleBlochl93}

\section{Experimental details}\label{sec:exp}
Single crystals of n-type non-intentionally doped $\beta$-Ga$_2$O$_3$ have
been purchased from a commercial supplier (Tamura, Japan). The sample thickness was 500 $\mu$m and the size
$4\times4$ mm$^2$ with the {\bf b}-axis normal to the sample plane. The samples have been irradiated at
room temperature with high energy electrons or protons to introduce intrinsic defects. Typical fluences
were 10$^{16}$ cm$^{-2}$. The irradiation conditions (12 MeV protons, 30 MeV electrons) were chosen so as  to guarantee a homogenous defect formation in the entire sample volume and so that no hydrogen is introduced in the sample due to
the irradiation. The EPR spectra were taken with an X-band
spectrometer (resonant frequency 9.3 GHz) and a variable
temperature (4K-300K) cryostat, which allowed in-situ optical excitation.
Angular variations of the EPR spectra were measured in three crystal planes. In addition to the
irradiation induced defect, the samples presented EPR spectra from a shallow donor and a weak
Fe$^{3+}$ spectrum,  which were discussed in our previous report.\cite{Jurgen18}

\section{Experimental results}\label{sec:exptresults}
\begin{figure}
  \includegraphics[width=8cm]{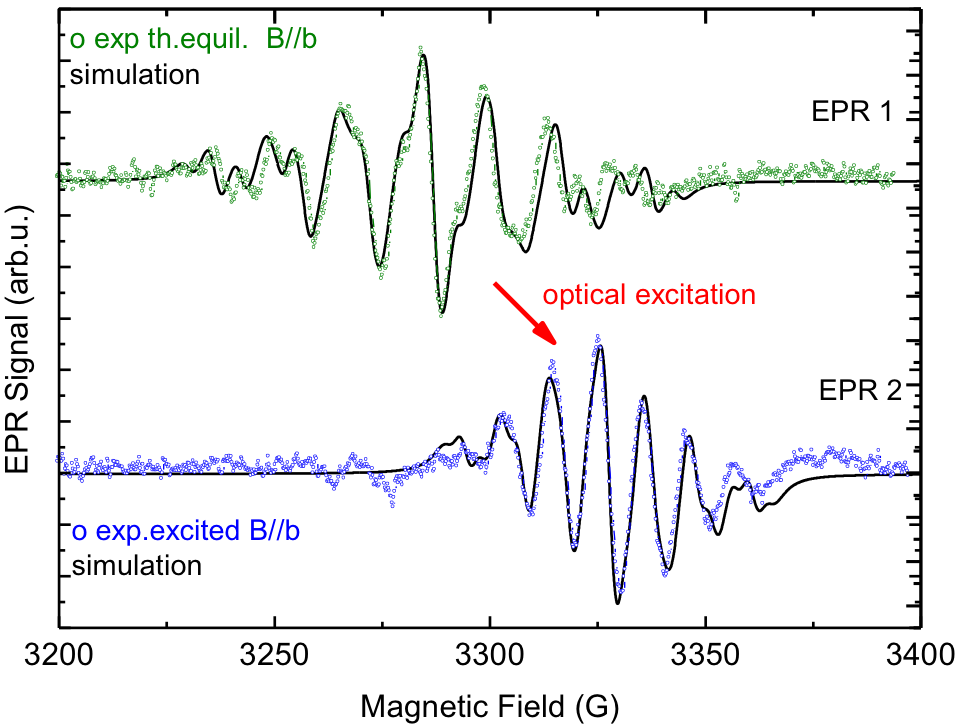}
  \caption{Experimental and simulated EPR spectra 
    in $\beta$-Ga$_2$O$_3$ for ${\bf B}\parallel{\bf b}$
    before photoexcitation (green) at $T=300$ K and 
    after photoexcitation (blue) at $T=52$ K. \label{figeprspec}}
\end{figure}

In Fig. \ref{figeprspec} we show the EPR spectra of the irradiation induced defect center for the applied magnetic field oriented ${\bf B}\parallel{\bf b}$ both before and after photoexcitation. 
We call these spectra respectively EPR1 and EPR2.
EPR1 occurs in the dark already at room temperature.
It displays a well-resolved multiplet structure
which can be simulated by a spin $S=1/2$ center, interacting with two equivalent Ga neighbors.
Due to the presence of two Ga isotopes ($^{69}\mathrm{Ga}$, $^{71}\mathrm{Ga}$) both with nuclear spin
$I=3/2$ but different isotopic abundances (60.1\%/39.9\%) and different nuclear moments in
these non isotopically modified samples, the hyperfine interaction gives rise to a characteristic
lineshape structure (the simulation is shown in Fig. \ref{figeprspec} as the black solid line). 
It clearly agrees with the spectrum reported previously by Kananen \etal\cite{Kananen17}

\begin{figure}[h]
  \includegraphics[width=8cm]{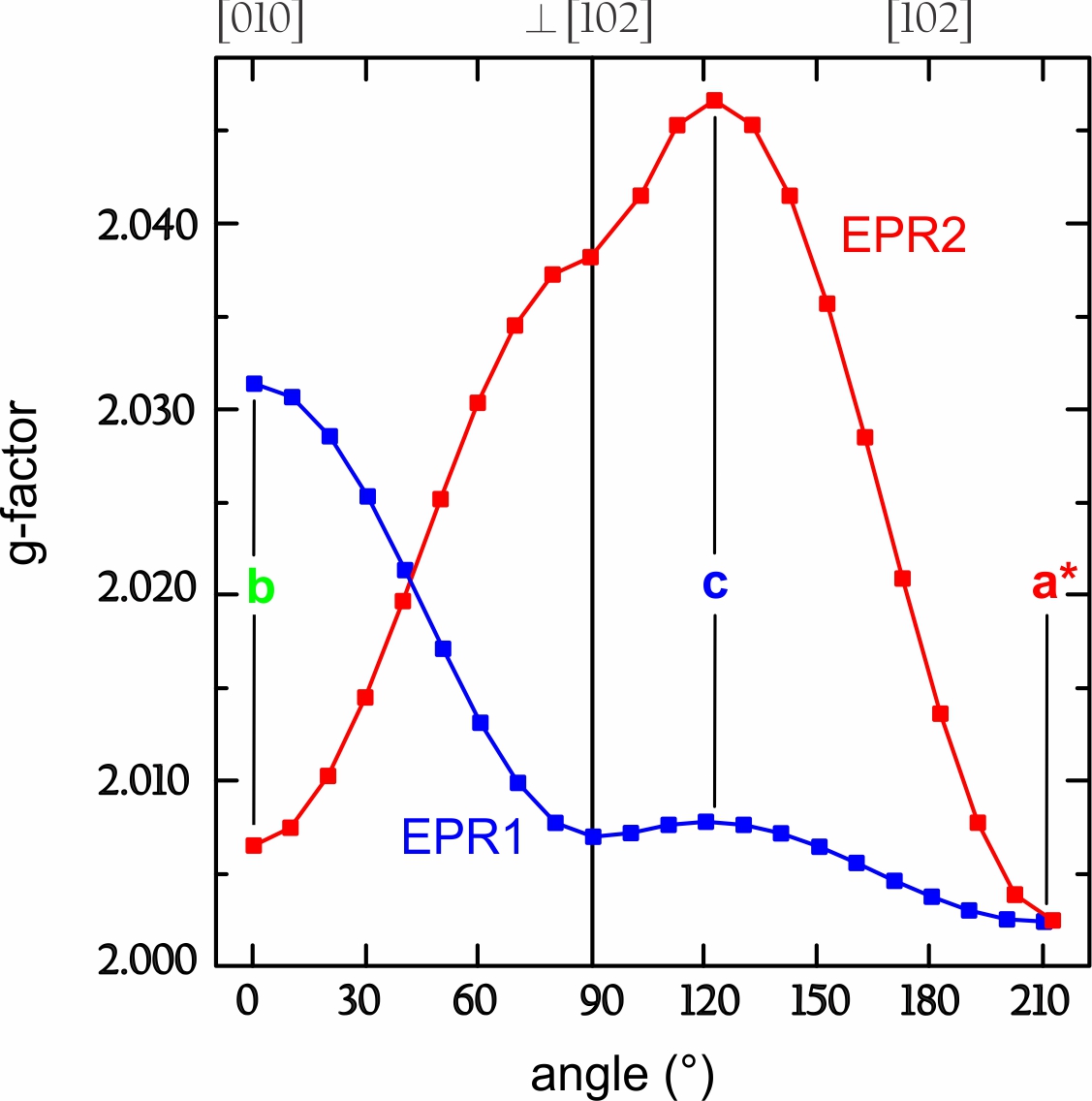}
%  \subfigure[]{\includegraphics[width=8cm]{angle1}}
%   \subfigure[]{\includegraphics[width=8cm]{angle2}}
  \caption{Angular variation of the $g$-factor for the variation of the magnetic
    field in two crystal planes for EPR1 (before) and EPR2 (after) photoexcitation. 
    \label{figangledep}}
\end{figure}

\begin{figure}[h]
  \includegraphics[width=9cm]{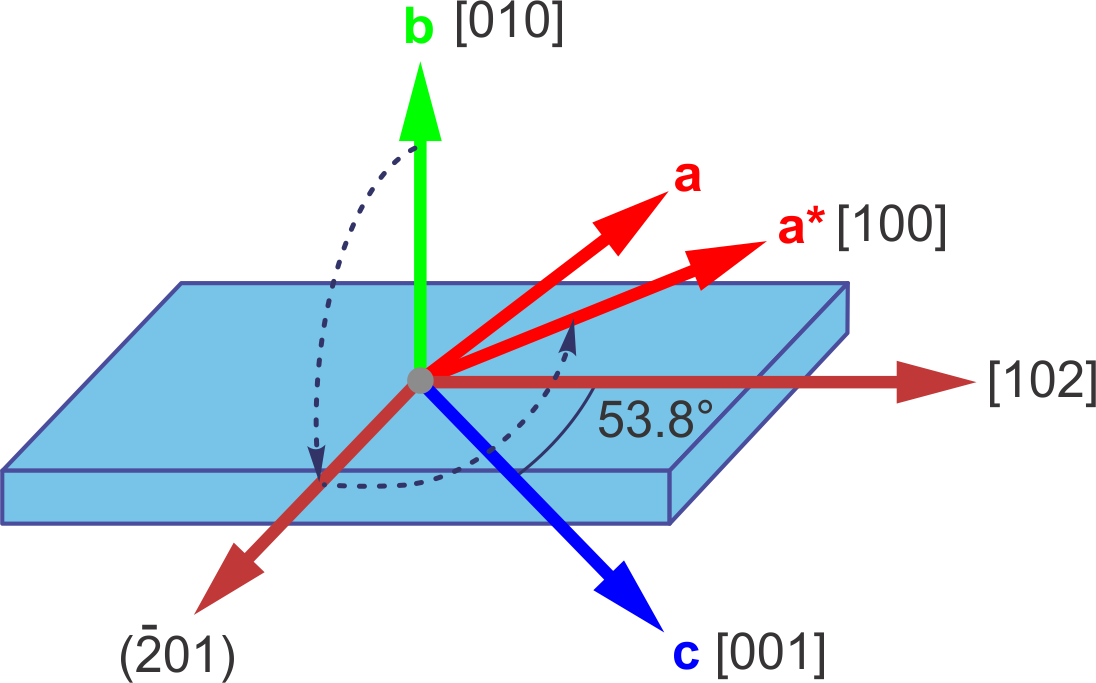}
  \caption{Sample indicating crystal orientations. [102] means
    ${\bf a}+2{\bf c}$, while (-201) means $-2{\bf a}^*+{\bf c}^*$,
    where the starred vectors are reciprocal lattice vectors, so that
    (-201) is perpendicular to [102]. The dashed curved lines indicate the
    angular variation of the magnetic field in two planes used in
    Fig. \ref{figangledep}. The angle between {\bf a} and {\bf c} is 103.7$^\circ$.
    \label{sample}}
\end{figure}

When the sample is photoexcited at low temperature ($T<100$ K) the spectrum EPR1 is erased and
replaced by a different spectrum, EPR2, which is metastable up to a temperature of $T=100$ K, at which
the EPR1 spectrum is regenerated. 
The angular variation of EPR1 and EPR2 in two planes is shown in Fig.\ref{figangledep}. 
Fig.\ref{sample} shows the crystallographic directions relative to the sample.
The main parameters extracted from the fit are summarized in Table \ref{tabepr}. 
The spectral dependence of the photoexcitation process was shown in Fig. 8 of Ref. \onlinecite{Jurgen18}
and shows the transition occurs rather abruptly for photon energies exceeding 2.8 eV.

%\begin{figure}[h]
%  \includegraphics[width=7cm]{figphotenergy}
%  \caption{Photoexcitation spectral dependence; EPR signal intensity
%    as function of photon energy. 
%    \label{figthreshold}}
%\end{figure}

The particular $g$-values and the superhyperfine interaction with two equivalent Ga neighbors
indicate for both centers a defect localized on a single oxygen site with two nearest Ga neighbors and are thus
most likely associated with Ga vacancy defects, which will be further substantiated by means of the calculations. 
The single spin $S=1/2$ nature of the spectrum indicates the $V_\mathrm{Ga}$ is in the  charge state $q=-2$.
The various one electron levels in the
gap closely above the valence band maximum (VBM) would be completely
filled in the $q=-3$ charge state, so that the $q=-2$ charge state has a
single hole.
The O hyperfine is not seen because the $^{16}\mathrm{O}$
isotope has no nuclear spin and is more than 99.9 \% abundant. The $^{17}\mathrm{O}$
isotope with a spin of $I=5/2$ has a natural abundance of only $3.8\times10^{-4}$. The hyperfine splitting is thus really a superhyperfine splitting because
it does not occur on the atom where the spin is dominantly localized but
on its neighbors.

\begin{table}
  \caption{EPR parameters $g$-tensor and hyperfine $A$ tensor for $^{69}$Ga;
    ${\bf b}$, ${\bf c}$ are the axes of the conventional unit cell, ${\bf a}^*$ is the reciprocal lattice vector in the ${\bf b}$-plane at 90$^\circ$ from the ${\bf c}$-axis. These
    are close to the principal axes of the $g$-tensor. 
     The HFI parameters for the $^{71}$Ga are obtained
     by multiplying by the ratio of their gyromagnetic factors which is 1.27059. The zero field
     splitting parameter $D$ corresponds to the corresponding $S=1$ spectra. 
         \label{tabepr}}
  \begin{ruledtabular}
    \begin{tabular}{lc}
      \multicolumn{2}{c}{Dark} \\ \hline
      $g_b$ & 2.0313 \\
      $g_c$ & 2.0079 \\
      $g_{a^*}$ & 2.0025 \\
      $A_b$ (G)  & 13.8  \\
      $A_c$ (G)  & 14.6 \\
      $A_{a*}$ (G) & 12.8 \\
      $D$ (MHz) & 250 \\
            \multicolumn{2}{c}{Photoexcited} \\ \hline
      $g_c$ & 2.0464 \\
      $g_{a^*}$ & 2.0024 \\
      $g_b$    & 2.0064 \\
            $A_b$ (G) & 9.8  \\
            $A_c$ (G) & 9.4 \\
            $A_{a*}$ (G) & 9.0 \\
            $D$ (MHz) & 322 \\
    \end{tabular}
  \end{ruledtabular}
\end{table}

\begin{figure}
   \subfigure[]{\includegraphics[width=8cm]{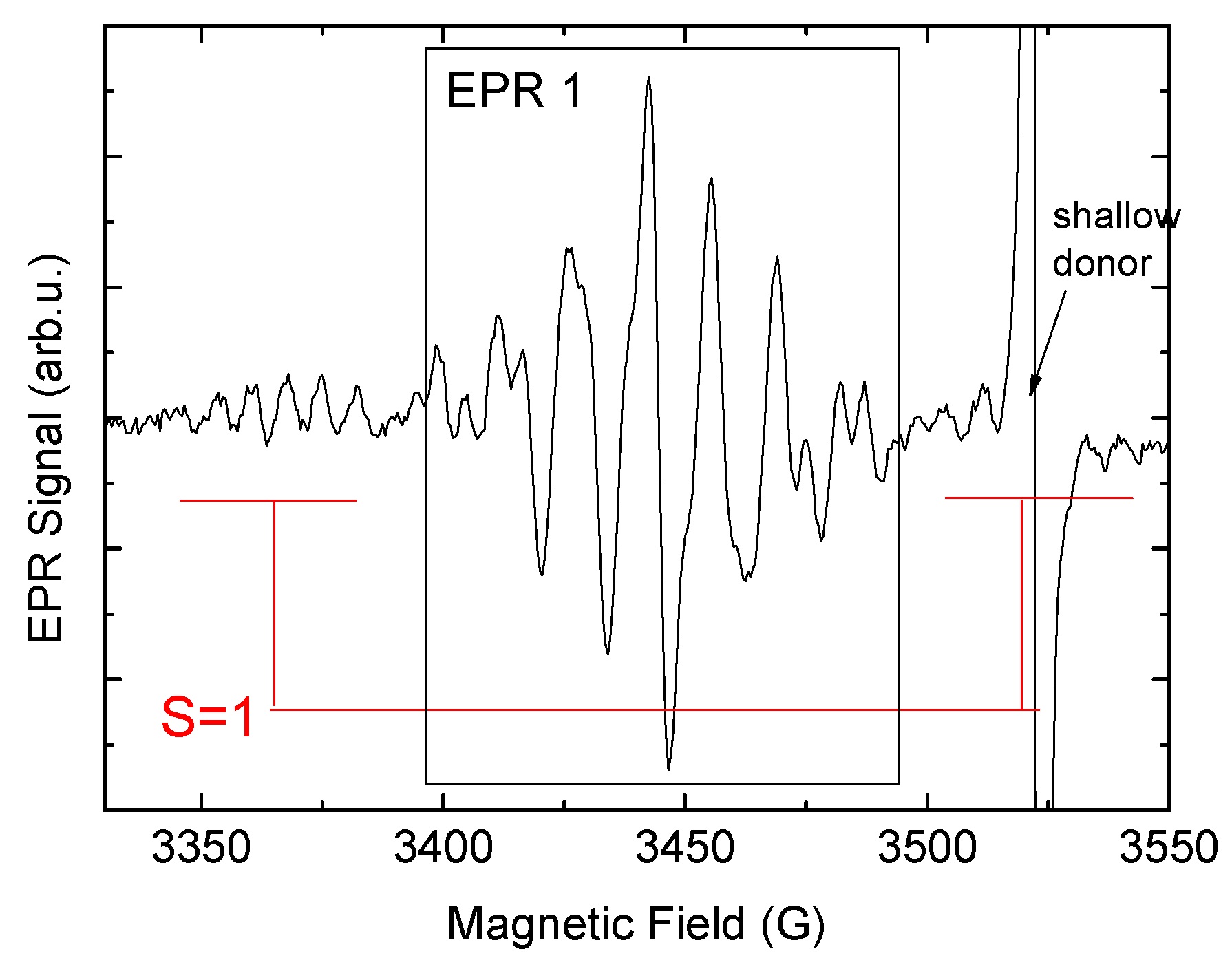}}
\subfigure[]{\includegraphics[width=8cm]{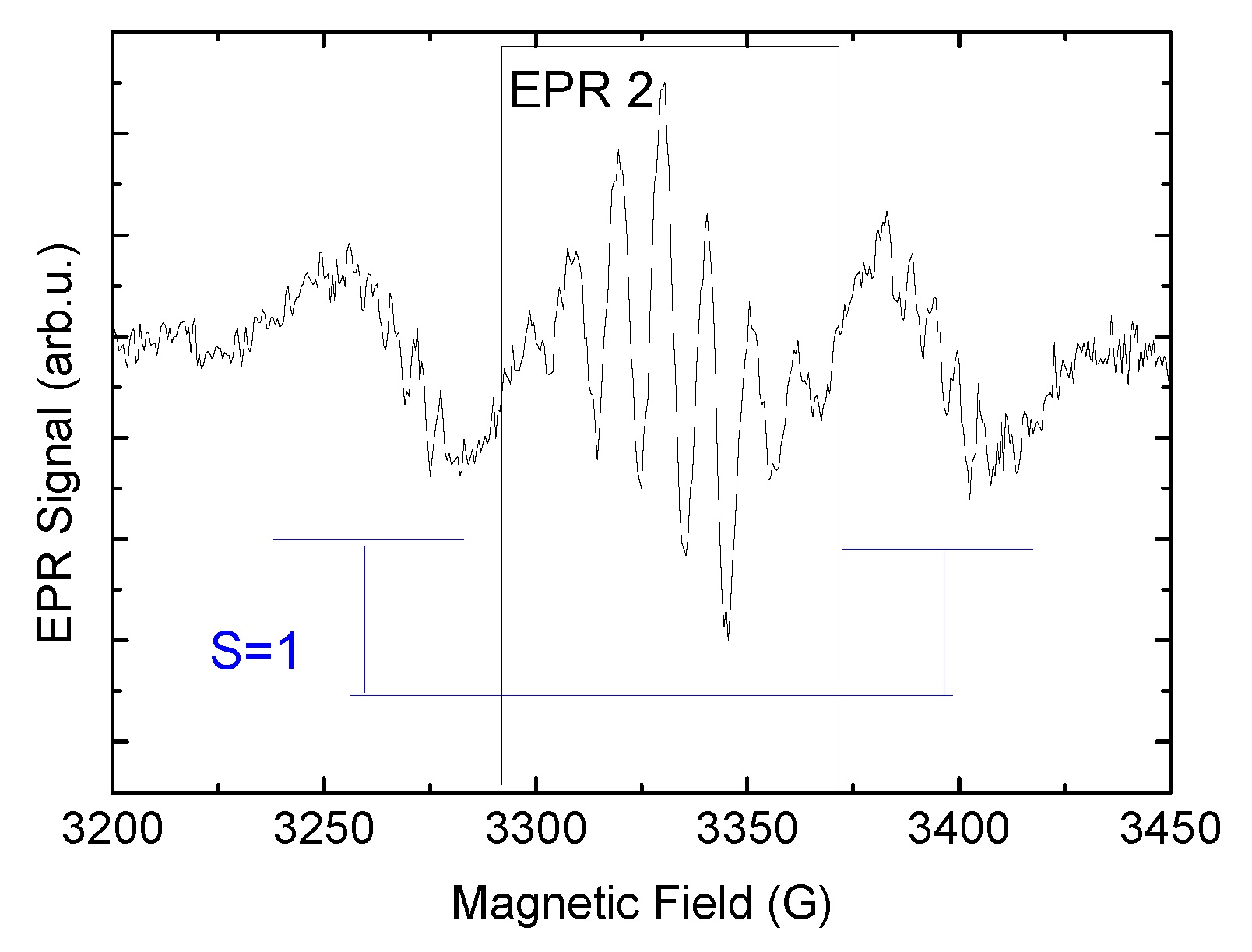}}
\caption{$S=1$ spectra accompanying the $S=1/2$ (a) before photoexcitation
  for ${\bf B}\perp{\bf b}$ and (b) after photoexcitation for
  ${\bf B}\parallel{\bf b}$.
  \label{figs1}}
\end{figure}

Each of the $S=1/2$ spectra reported above
is accompanied by a corresponding distinct $S=1$ spectrum (see Fig. \ref{figs1}). While the first  $S=1$ spectrum
related to EPR1 was also previously reported by
Kananen\cite{Kananen17} we here emphasize that there are two distinct
$S=1$ spectra. Each has a $g$-tensor equal to the corresponding $S=1/2$
and a hyperfine splitting with about half the value of the contact hyperfine
interaction for the $S=1/2$. This indicates it consists of  two weakly interacting $S=1/2$
spins. It means that the same defect can exist in two charge states with either one or two holes.
The spin Hamiltonian for the $S=1$ case contains a term ${\bf S}\cdot{\bf D}\cdot{\bf S}$
which is  usually written as
\begin{equation}
  H=D[S_z^2-S(S+1)]+E(S_x^2-S_y^2)
\end{equation}
with $D=\frac{3}{2}D_z$ and $E=\frac{1}{2}(D_x-D_y)$. 
Here, the zero-field splittings are nearly axial with $E=0$. Their $D$ values are given in
Table \ref{tabepr}. They have principal axes at 20$^\circ$ from the [102] axis
toward the ${\bf c}$-axis in the {\bf b}-plane for EPR1 and
parallel to [102] for EPR2.  

\section{Computational models}\label{sec:models}
Under irradiation both vacancies
and interstitials of both Ga and O  atoms are likely to occur. However,
we focus on Ga-vacancies and O-interstitials because only these are expected
to have spin localized on an O atom, as is clearly established by the experiment.

The first models are: (M1) $V_\mathrm{Ga1}$ with
spin localized on O$_{(1)}$ neighbor, (M2) $V_\mathrm{Ga2}$ with
the spin localized on O$_{(2)}$ on the mirror-plane, (M3)
a $V_\mathrm{Ga2}$ with tilted spin
localized on two O$_{(1)}$ on either side of the mirror plane. We will
also refer to this as the ``tilted spin'' model.  This model is the
structure obtained in Ref. \onlinecite{Deak17} as ground state
for the $V_\mathrm{Ga2}$.  The M2 model with
the spin on the mirror plane O$_{(2)}$ can be stabilized when using 
a different initial displacement of the atoms. In hybrid functional calculations
(with the same parameters as in Ref. \onlinecite{Deak17}) that model
lies 0.27 eV higher in energy. 
% UG: if we agree in prefering the hybrid functional result we should leave 
%     out the followinh sentence.
%However, in a larger cell  (240 atoms) and using GGA+U it was found to have lower energy than M3.
%Although it is thus not yet entirely clear which of the two is the actual
%ground state, we prefer the hybrid functional result for the purpose of
%obtaining energetic ordering.  
Nonetheless we will also consider M2
for the purposes of comparing its EPR parameters with experiment, in particular because
it corresponds to the model proposed by Kananen \etal\cite{Kananen17} for EPR1. 

As for model M3, it should
be pointed out that in this model the wave function tends to become asymmetric
with spin localized on one O$_{(1)}$ on one side of the mirror plane.
However, two variants exist:
a left, and a right one, and the experiment which averages macroscopically
over many centers in the sample, would then see an average of the two.
Alternatively, the spin may dynamically flip back and forth between the two sides in the same defect
on the time scale of the experiment. This would correspond to a dynamic Jahn-Teller distortion.  
As a convenient
way to calculate the $g$-tensor for this titled spin model,  we have enforced the symmetry in the calculation
with the single spin distributed equally over the two sides. While this
would then seem to entail hyperfine interaction with 4 instead of 2 Ga,
or rather two slightly inequivalent pairs, we should keep in mind that
the spin is not really spread over the two sides but temporarily on one or the other in a dynamic sense. Thus it still only interacts with two Ga at any given moment. 
 Please note that either of the two approaches, averaging the
left and right variants of the unsymmetrized model M3 or the symmetrized M3
will give the same $g$-tensor.

\begin{table}
  \caption{Total energy differences (in eV) of Ga vacancies and their stoichiometrically equivalent complexes in different charge states
    relative to the $q=0$ state of $V_\mathrm{Ga1}$. These are calculated
    using the hybrid functional.\label{tabtote}}
  \begin{ruledtabular}
    \begin{tabular}{lcccc}
q      	& 0	& -1	& -2	& -3 \\ \hline
$V_\mathrm{Ga1}$ &	0	& 4.62	& 9.96	& 16.25 \\
$V_\mathrm{Ga2}$ &	0.68	& 4.77	& 9.84	& 15.67 \\
$V_\mathrm{Ga1}-\mathrm{Ga}_{ic}-V_\mathrm{Ga1}$ &	0.99 &	4.91 &	9.70 &	15.10 \\
$V_\mathrm{Ga1}-\mathrm{Ga}_{ib}-V_\mathrm{Ga1}$ &	0.33 &	3.89 &	9.38 &	15.54\\
    \end{tabular}
  \end{ruledtabular}
\end{table}

Next, we consider complexes
such as the $V_\mathrm{Ga1}-\mathrm{Ga}_{ib}-V_\mathrm{Ga1}$
complex, which we label M4. We will refer to it as the (b)-complex for short. 
Figures of the models including the results of the spin density
appear in the next section. 
These type of complexes were found to have lower energy
than a simple $V_{\rm Ga1}$
by Varley \etal \cite{Varley11} in their most negative charge state
and could occur
by means of a Ga diffusion process.\cite{Kyrtsos17} There are actually
two nonequivalent such complexes, as defined in Fig. 4 of
Kyrtsos \etal\cite{Kyrtsos17}. The M4 model corresponds to the one
indicated as the (b)  location of the interstitial Ga. The two O$_{(1)}$
on which the spins localizes and which are related
by a center of inversion at the (b)-interstitial site
in this case have each two Ga$_{(2)}$ neighbors.
In the other case, indicated by (c) in Fig. 4 of Ref. \onlinecite{Kyrtsos17}
the spins localize on O$_{(3)}$ type atoms which have each three Ga neighbors.
Our total energy calculations using the hybrid functional \cite{Deak17} (shown in Table \ref{tabtote})
indicate that this (c)-complex has the lowest energy among the complexes for the $q=-3$ charge.
However, for the EPR relevant charge states $q=-2$ and $q=-1$ the (b)-complex
has the lowest energy. We will later discuss their likelihood of being in the
$q=-2$ or $q=-1$ charge states which are relevant as EPR active states. 
We will refer to the (c)-complex as the M5 model. 
Another complex consists of $V_\mathrm{Ga1}-\mathrm{Ga}_{ia}-V_\mathrm{Ga2}$
is called the M6 model. 

We found for the M4  complex that the spin stays localized on one O$_{(1)}$
similar to the simple $V_\mathrm{Ga1}$ model and as a result exhibits slightly
different relaxation near each $V_\mathrm{Ga1}$ in the model. Obviously
there are then two variants but their $g$-tensors are exactly the same because
they are related by an inversion symmetry. Thus no averaging is required.
These models were initially relaxed with the DFT$+U$ method but later checked
with hybrid functional when their transition levels were determined. 
We note, that similar to M3 we could here also have symmetrized the model
explicitly, but we find that this makes little difference and hence we
do not show the results for the symmetrized version here.

The above models focused on Ga-vacancies. However, in a previous work,\cite{Kananensth}
the self-trapped hole was proposed as a model for EPR2. 
We consider two self-trapped hole (STH) models  with the hole trapped either on
O$_{(1)}$\cite{Varley12,Deak17} (M7) or on two O$_{(2)}$ (M8).\cite{Deak17}
The site on which the hole is trapped is determined by the initial
displacement of the neighboring atoms from which the system converges
to the closest
local minimum in the energy. So, there can indeed be different hole trapping
sites and hopping between them could occur at higher temperature. 
The M8 model has two Ga$_{(2)}$ and one Ga$_{(1)}$ neighbor. If the Ga$_{(1)}$
were 
far enough removed due to the relaxations around the hole, then it could be 
a plausible model for the EPR centers at hand. However, we will show
that this is not the case and 
that the hyperfine on the third Ga is actually higher than the first two. 
The M8 model has
already spin distributed over two O$_{(2)}$ atoms and is thus likely
to cause hyperfine on more Ga. We find below that this is indeed the case
and this rules out that model.  Self-trapping on O$_{(3)}$
was not found to occur in previous studies.\cite{Deak17}

Another type of defect which is expected to have spin density localized on oxygen
is interstitial oxygen, O$_i$. The latter adopts a split interstitial configuration.
It consists of an O$_2$ dumbbell on a lattice O-site.
Among such models,
the O$_2$ located on the O$_{(1)}$ site is the most promising. In fact, in the
neutral charge state, it was found to be oriented close to the {\bf c}
direction. A metastable state occurs when a hole is added to this
defect in this configuration.
We label this model as M9. We should mention
that the model for this metastable state was obtained here using the DFT+U approach 
rather than the hybrid functional.  However, a previous work\cite{Deak17} at the 
hybrid functional level also found a split interstitial O$_i$; the transition levels 
of which will be discussed later in sec. VII, A. Eventually, its ground state has the
dumbbell rotated closer to the {\bf b} direction. However, the {\bf c}-oriented
dumbbell with  single hole spin located on it was here considered  {\sl a-priori} 
as an attractive model  to explain  EPR2 because the spin might become more localized near
one of the two O and thereby have reduced hyperfine interaction with the
third Ga, which is a Ga$_{(1)}$ and because it was likely to have its
$g$-tensor along {\bf c}.  As a final related system, we
consider the $\mathrm{O}_i-V_\mathrm{Ga1}$ complex in which that Ga$_{(1)}$
is removed. This is labeled the M10 model. It was relaxed within DFT$+U$. 

\section{Computational results}\label{sec:compresults}
\begin{table}
  \caption{Calculated EPR parameters for various Ga-vacancy related models
    (labeled M1 to M6), the O on which the spin is localized in column 2. 
    The $g$-tensor is given by $\Delta g_i$ (difference
    from the free electron $g$-value) and principal values ordered from high to low;
    the corresponding axes are specified by their polar angle $\theta_i$ 
    from the $\hat{\bf z}={\bf b}$ axis and azimuthal angle
    $\phi_i$ from the $\hat{\bf x}={\bf a}^*$ axis, both in $^\circ$, followed
    by which crystal axes are closest to these directions. 
    For the hyperfine interaction
    HFI, the number of Ga atoms with strong HFI and the 
    isotropic component of the $A$ tensor (in $10^{-4}$T=Gauss) is given for 
    the $^{69}$Ga isotope. Note that these HFI are from PBE$+U$ wave functions 
    with $U=4$ eV; for further discussion of the $U$ dependence see text.
    For comparison the corresponding data of the experimentally observed $S=1/2$ 
    spectra is also given (first two lines). 
    \label{tabcalceprvga}
   }
  \begin{ruledtabular}
    \begin{tabular}{l|c|ccc|cc}
      Model & O&\multicolumn{3}{c|}{$g$-tensor} & \multicolumn{2}{c}{HFI} \\\hline
      $\begin{array}{cc}\mathrm{label}\\\mathrm{structure}
      \end{array}$
      &type& \multicolumn{3}{c|}{$\begin{array}{ccc}\Delta g_1&\Delta g_2& \Delta g_3\\\theta_1&\theta_2&\theta_3\\
          \phi_1&\phi_2&\phi_3 
        \end{array}$} & $\#$Ga & $A$ (G) \\ \hline
      EPR1  &       &  0.0289 & 0.0056 & 0.0002 & 2 & 13.7 \\
      &       &  $b$    & $c$    & $a_*$  & & \\ \hline
      EPR2  &       &  0.0441  & 0.0041 & 0.0001 & 2 & 9.4 \\
            &       & $c$      & $b$    & $a_*$  & & \\ \hline
      M1    & O$_{(1)}$ & 0.0219 & 0.0175 & 0.0045 & 2 & $-22$\\
      $V_\mathrm{Ga1}$ && 0  & 90 & 90 & & \\
      &    && 22 & -68 & & \\
        && $b$ & $a_*$ & $c$ & & \\ \hline
%      M1 &O$_{(1)}$&  0.0228 & 0.0187 & 0.0051   & & \\
%      $V_\mathrm{Ga1}$ && 7   & 83 & 89 \\
%      240 && 19        & 26 & $-64$      \\
%      && $b$ & $a_*$ & $c$ & & \\\hline
      M2 & O$_{(2)}$ &0.0235 & 0.0161& 0.0062  & 2 & $-22$ \\
      $V_\mathrm{Ga2}$&& 0  & 90 & 90 & & \\
       &&  & -22 & $68$ &  &  \\
      && $b$ & $a_*$ & $c$ & & \\\hline
%       M2 &O$_{(2)}$& 0.0202 & 0.0180& 0.0089  &  &  \\
%      $V_\mathrm{Ga2}$&& 4  & 86 & 86 & & \\
%       240&& 58 & -47 & $43$ &  &  \\
%       && $b$ & $a_*$ & $c$ & & \\ \hline
       M3 & O$_{(1)}$&0.0349 & 0.0180 & 0.0160 & 2 & $-21$  \\
$V_\mathrm{Ga2}$       && 88     & 80 & 10 &  2 &  $-16$  \\
       && -69     & $20$ & $30$ &  & \\
       && $c$ & $a_*$ & $b$ & &  \\ \hline
       M4 & O$_{(1)}$ & 0.0228 & 0.0124 & 0.0025 & 2 & $-21$ \\
       $V_\mathrm{Ga1}-\mathrm{Ga}_{ib}-V_\mathrm{Ga1}$ && 0 & 90 & 90 & &
        \\
          &&  & 70 & $-20$ & &  \\
       && $b$ & $c$ & $a_*$ & & \\ \hline
%       M4' &2O$_{(1)}$ & 0.0316 & 0.0140 & 0.0008 & 4 & $-16$\footnote{In PBE the $A$ was $-20.7$ and in $PBE+U$ it was impossible to keep the spins spread over the two
%         sides, but in M4, PBE  to PBE$+U$ gave a reduction of $A$ by a factor
 %        $20.7/26.7=0.775$. Hence we reduced the PBE (M4')value by this factor.} \\
 %      $V_\mathrm{Ga1}-\mathrm{Ga}_{ib}-V_\mathrm{Ga1}$ && 0 & 90 & 90 & & \\
 %      160                                           &&   & 73 & $-16$ & & \\
 %      && $b$ & $c$ & $a_*$ & & \\ \hline
       M5 &O$_{(3)}$ & 0.0187 & 0.0177 & 0.0036 & 2 & $-21$ \\
       $V_\mathrm{Ga1}-\mathrm{Ga}_{ic}-V_\mathrm{Ga1}$ &&0 & 90 & 90 & 1 & $-20$ \\
        &&  & $-20$ & 70 & & \\
       &&  $b$ & $a_*$ & $c$ & & \\ \hline
       M6 & O$_{(2)}$ & 0.0342 & 0.0125 & 0.0029 & 1 & $-27$ \\
       $V_\mathrm{Ga1}-\mathrm{Ga}_{ia}-V_\mathrm{Ga2}$ && 69 & 21 & 89 & 1 & $-20$ \\
        && $2$ & $4$ & -88 &   & \\
       &&     $a*$    & $b$   & $c$ && \\
    \end{tabular}
  \end{ruledtabular}
\end{table}

\begin{table}
  \caption{EPR parameters for self-trapped holes, oxygen interstitial
    and related models. The table is arranged similar to
    Table \ref{tabcalceprvga}. For further details see the respective caption.\label{tabcalceprO}}
  \begin{ruledtabular}
    \begin{tabular}{l|c|ccc|cc}
     Model & O&\multicolumn{3}{c|}{$g$-tensor} & \multicolumn{2}{c}{HFI} \\\hline
      $\begin{array}{cc}\mathrm{label}\\\mathrm{structure}
      \end{array}$
      &type& \multicolumn{3}{c|}{$\begin{array}{ccc}\Delta g_1&\Delta g_2& \Delta g_3\\\theta_1&\theta_2&\theta_3\\
          \phi_1&\phi_2&\phi_3 
        \end{array}$} & $\#$Ga & $A$ (G) \\ \hline
       M7 & $O_{(1)}$ &  0.0214 & 0.0205 & 0.0090 & 2 & $-8$ \\
       STH $h^+_{O_{(1)}}$ &&        90 & 0 & 90 & 1 & $-16$ \\
        &&        $-73$ & & 17 & & \\
       &&  $c$ & $b$ & $a_*$ & & \\ \hline
       M8 & 2O$_{(2)}$ & 0.0172 & 0.0112 & 0.0042 & 2 & $-13$ \\
       STH $h_{2O_{(2)}}^+$ && 0 & 90 & 90 &  2&  $-12$ \\
                   &&   & $2$  & -88 &  2 & $-9$ \\
       && $b$ & $a_*$ & $c$ & & \\ \hline
       M9 &O$_{(1)}$+O$_i$ & 0.0283 & 0.0037 & 0.0011 & 2 & $-8$ \\
       O$_i$ &&         90      & 0      & 90 & 1 & $-19$ \\
          &&         $79$      &        & -11  & & \\
       &&         $c$     & $b$    & $a_*$  & & \\ \hline
       M10 & O$_i$ & 0.0241 & 0.0051 & 0.0008 & 2 & $-3$ \\
       $\mathrm{O}_i-V_\mathrm{Ga1}$ && 90 & 0 & 90 & & \\
                                  && -84 & & $7$ & & \\
                                     && $c$ & $b$ & $a_*$ & & \\ 
    \end{tabular}  
  \end{ruledtabular}
\end{table}
                                    
The $g$-tensors principal values and axes as well as the HFI
splittings are summarized in Table \ref{tabcalceprvga}
for Ga-vacancy related models (M1-M6)  and in Table \ref{tabcalceprO}
for the self-trapped and interstitial O related models (M7-M10).

\begin{figure*}
  \includegraphics[width=16cm]{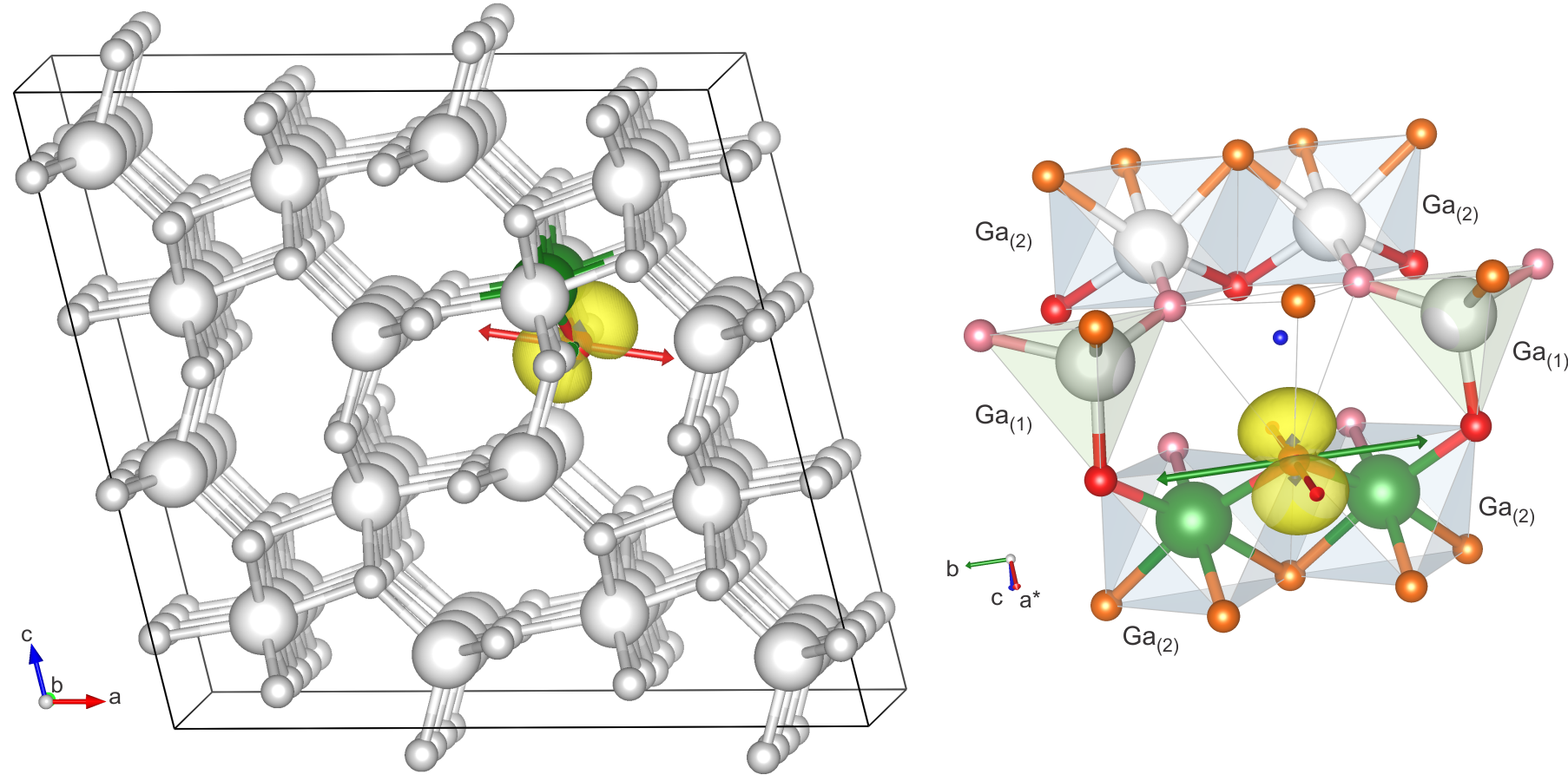}
  \caption{$V_\mathrm{Ga1}$ (M1) structure, spin density in yellow, $g$-tensor
    principal axes indicated by double arrows
    with length proportional to the $\Delta g$  (deviation from free electron
                 value $g_e = 2.002391$), green colored Ga atoms
    are the ones with strong HFI. On the left it is viewed in the
    supercell, on the right the local structure is viewed from
    a different angle, the small O
    spheres are color coded red O$_{(1)}$, pink O$_{(2)}$,
    orange O$_{(3)}$ and the polyhedra surrounding the Ga and their type
    are indicated. The tetrahedral vacancy $V_\mathrm{Ga1}$ is indicated as a colorless tetrahedron with a small blue sphere. 
  \label{figvga1}}
\end{figure*}

\begin{figure*}
  \includegraphics[width=16cm]{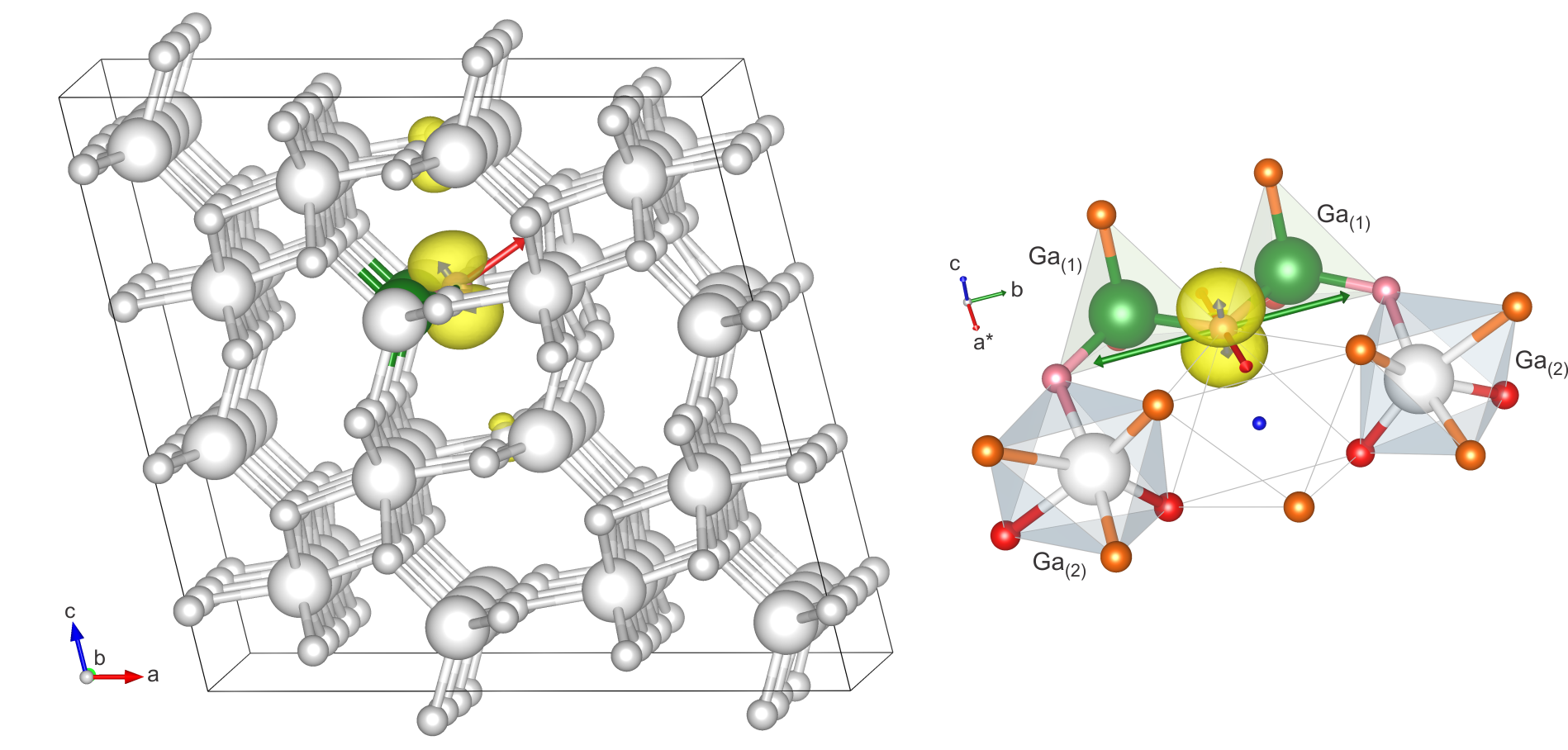}
  \caption{$V_\mathrm{Ga2}$ (M2) structure, spin density in yellow, $g$-tensor
   and  Ga atoms with strong HFI. See Fig. \ref{figvga1} for details.
  \label{figvga2}}
\end{figure*}

The spin density, $g$-tensor and atoms with large HFI
are shown
in two views for $V_\mathrm{Ga1}$ (M1) in Fig. \ref{figvga1} and
for  $V_\mathrm{Ga2}$ (M2) in Fig. \ref{figvga2}. The results for the
$V_\mathrm{Ga1}-\mathrm{Ga}_{ib}-V_\mathrm{Ga1}$ complex (M4) is shown
in Fig. \ref{figM4} while those of M5, M6 are shown in Supplementary
Information.\cite{supinfo}

\begin{figure*}
  \includegraphics[width=16cm]{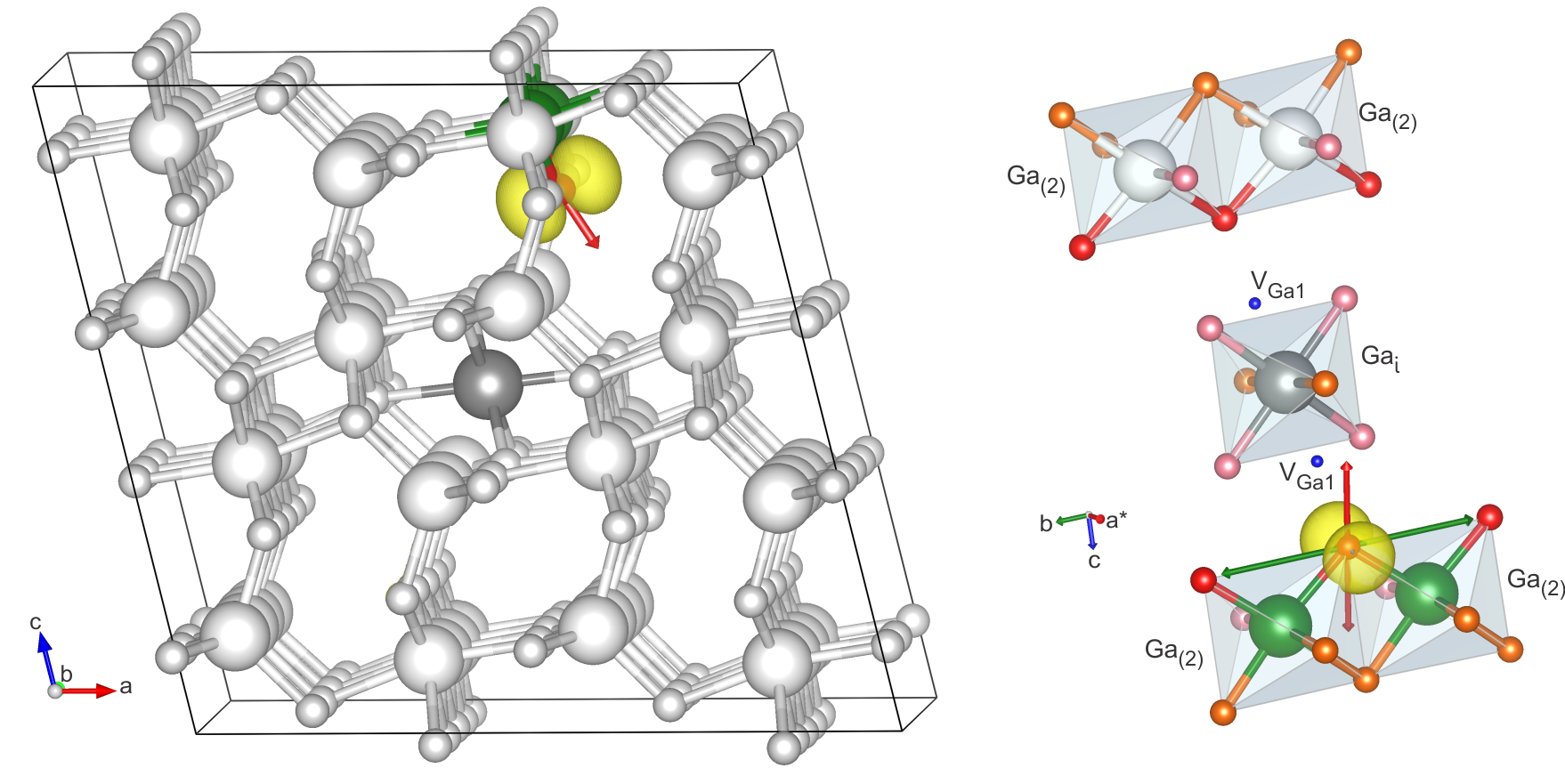}
    \caption{$V_\mathrm{Ga1}-\mathrm{Ga}_{ib}-V_\mathrm{Ga1}$ (M4) structure, spin density in yellow, $g$-tensor
      principal axes (green is larger component, red intermediate) and Ga atoms
      with strong HFI. Details as in Fig. \ref{figvga1}.
  \label{figM4}}
\end{figure*}

\section{Discussion}\label{sec:discussion}
\subsection{Models for EPR1}
We now compare the calculated results with the experimental results.
Starting with the simple $V_\mathrm{Ga}$ models, we first note that
both $V_\mathrm{Ga1}$ and $V_\mathrm{Ga2}$  in the M2 model, 
have the main $g$-tensor component along the {\bf b} direction in
agreement with the experimental data for EPR1.  
The $g$-tensor however differs
from the experimental one in two respects: first, the calculated one
has a large, an intermediate and a small principal values, whereas the
experimental one has only one large $\Delta g$ and two close significantly
smaller values. Second, the smallest $g$ principal value direction in
the calculation is close to ${\bf c}$ (within about 20$^\circ$), while in
the experiment it is along ${\bf a}^*$.

The direction of the smallest
$g$-component can be seen to correspond to the direction of the $p$-like
spin density (see \eg Fig. \ref{figvga2}). This in fact makes sense
from a perturbation theory point of view and in the empirical
model used in Kananen \etal\cite{Kananen17}. In a simple perturbation
theory picture, the $\Delta g$ arises from cross-terms between the
orbital Zeeman ${\bf B}\cdot{\bf L}$ and spin-orbit coupling
${\bf S}\cdot{\bf L}$ terms, 
which lead to an effective ${\bf B}\cdot{\bf S}$ term and hence a change
in the gyromagnetic factor in the spin Zeeman term,
\begin{equation}
  \Delta g_{ij}\propto\sum_n\frac{\langle0|L_i|n\rangle\langle n|L_j|0\rangle}{E_n-E_0},
\end{equation}
where $|0\rangle$ is the one-electron state with
unpaired spin and $|n\rangle$ all other states. 
If we pick the
direction of the spin density as the $p$-orbital momentum quantization axis $z$,
then the angular momentum matrix elements involving $L_z$ are zero and give no
contribution for the $g_{zz}$, \ie the deviation from
       the free electron value $g_e \approx 2.002319$, in that direction
whereas in the orthogonal directions,
matrix elements of $L_+$ and $L_-$ enter and can give a non-zero $\Delta g$.
Furthermore, we can see that the highest $\Delta g$ occurs along the
direction of the line that joins the two Ga atoms with strongest
HFI.  This indicates that the O $p$-orbital spread in this direction
contributes more to close lying excited states. Of course, this is only a
rough guide because the eigenstates of the defect are not purely O $p$-states.

In spite of the shortcomings in the comparison of theory and experiment
mentioned above and considering the uncertainties in
$\Delta g$ and its principal directions for the smaller $\Delta g$ components,
it is remarkable that both types of Ga vacancies give very similar
results and in reasonable agreement with the major features of the experiment.
This already strongly indicates a Ga-vacancy as basic model for EPR1. 
We will further refine the model below but first briefly discuss the hyperfine
splitting for these models. 

The hyperfine splitting for
both of these models is on the two equivalent Ga atoms that are connected to the
spin-carrying O. The fact that
the hyperfine is on two Ga neighbors agrees with experiment but its value
depends strongly on the functional used as already mentioned in Sec.\ref{sec:method}.
Within pure PBE-GGA, its value is significantly
overestimated (about $-29$ G). With hybrid functional it is even larger, namely $-32$ G.
Within DFT$+U$, the value depends strongly on $U$. For $U=4,8,10$ eV its values
are respectively $-22,-17,-16$ G. So, they become smaller with increasing $U$,
but even for $U=10$ eV they are still somewhat overestimated. A value of $U>10$ eV
appears unreasonable, so this indicates that the $+U$ approach is still imperfect
to describe the spread and shape of the wave function or that other effects not included
at this point, such as dynamic Jahn-Teller effects or correlation effects not captured by
DFT are required.  The other possibility is, of course, that we have not yet found
the correct model, but, as can be seen from Table \ref{tabcalceprvga} all vacancy related models have hyperfine
interactions of the same order of magnitude. 

Next, we consider the various Ga-vacancy-pair Ga-interstitial complexes
(models M4-M6, see also Fig. \ref{figM4} and Supplementary Information\cite{supinfo}).
Among these, it is clear that only M4 
gives reasonable agreement with experiment. M5 has hyperfine interaction
with three Ga and its $g$-tensor has two almost equal large values
0.018 along ${\bf b}$ and ${\bf a}^*$ in disagreement with
both EPR1 and EPR2. M6 has its largest value along ${\bf a}^*$,
the spin spread over two equivalent O$_{(2)}$ and significantly
different hyperfine on two pairs of Ga.
In contrast, M4 (Fig. \ref{figM4}) has a $g$-tensor maximum value along ${\bf b}$,
the next along ${\bf c}$ and the smallest along ${\bf a}^*$, in agreement
with the experiment. Furthermore, we may consider this as the actual ground
state of the $V_\mathrm{Ga1}$, since it has lower energy than a simple $V_\mathrm{Ga1}$
and is formed from the latter by a simple migration of a lattice Ga toward
and interstitial position.  
Finally, in this model, as mentioned in Sec. \ref{sec:method} we may have to 
consider a dynamic Jahn-Teller situation where the symmetry breaking occurs dynamically
and the spin flips back and forth between the two sides of the defect complex,
which may introduced a  hyperfine reduction factor. This would
help to correct the HFI overestimate for this model but not
for the single vacancies.

\begin{figure}
  \includegraphics[width=8cm]{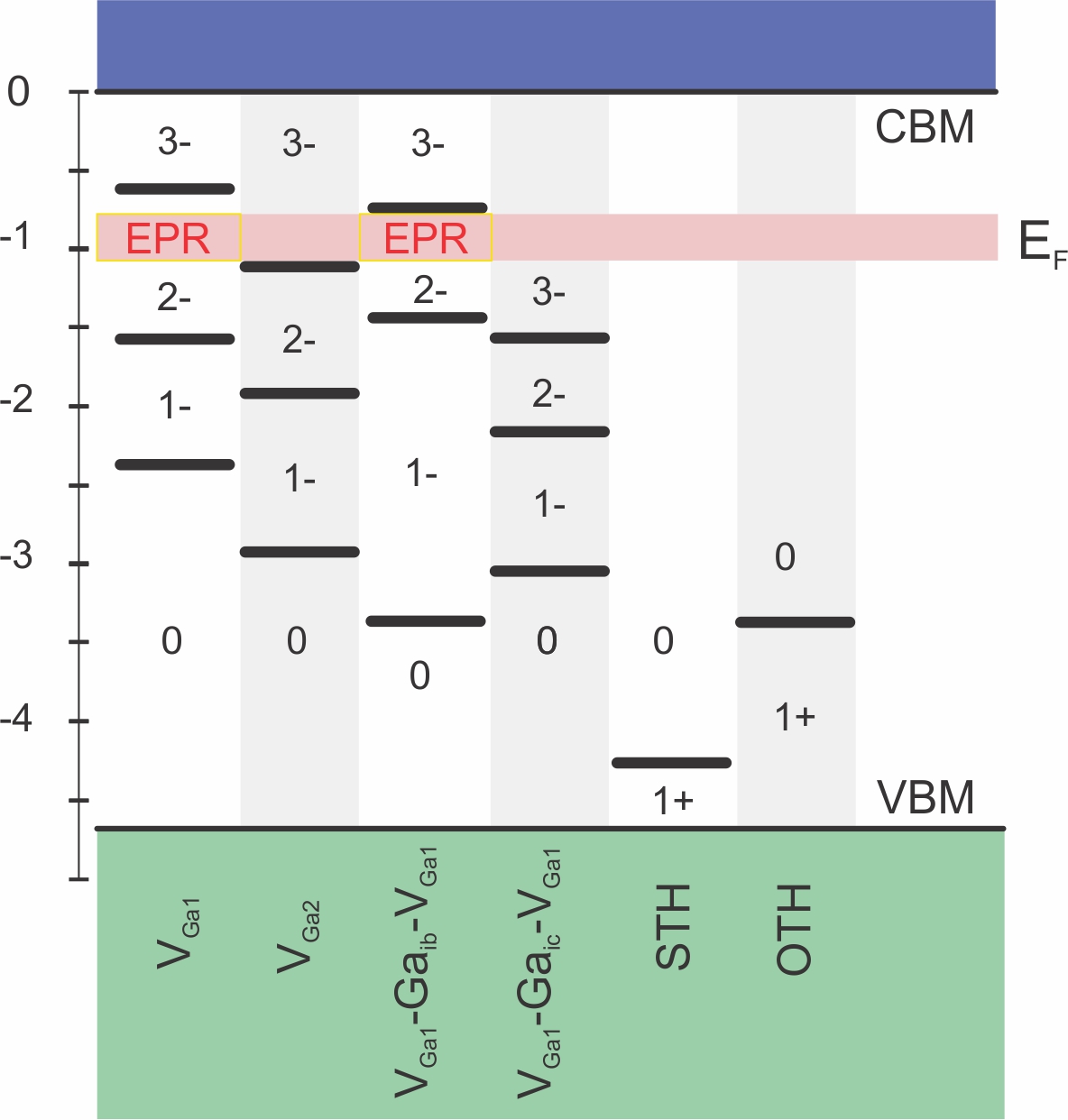}
  \caption{Transition levels in the gap and expected Fermi level range (in eV)
    after irradiation, based on which $V_\mathrm{Ga}$ related centers are EPR active. 
    The $V_\mathrm{Ga2}$ level belongs to ther tilted spin model M3. 
    \label{figtranslev}}
\end{figure}

\begin{table}
  \caption{Transition levels of various defects relative to the conduction band minimum (in eV) and calculated using the
    hybrid functional.\label{tabtranslev}}
  \begin{ruledtabular}
    \begin{tabular}{l|dddd} \\ 
      &\multicolumn{1}{c}{$V_\mathrm{Ga1}$} & \multicolumn{1}{c}{$V_\mathrm{Ga2}$} & \multicolumn{1}{c}{(b)-complex} &  \multicolumn{1}{c}{(c)-complex} \\ \hline
2-/3- 	& 	-0.67 &	-1.16 & -0.74 &  -1.59 \\
1-/2-	& 	-1.63&	-1.91 &	-1.42 & -2.20 \\
0/1-	&        -2.35 & -2.90 & -3.34 & -3.06 \\ 
    \end{tabular}
  \end{ruledtabular}
\end{table}

Our conclusion thus far is that the M4 double vacancy-interstitial complex is the best fitting
model for EPR1. However, one expects that during irradiation vacancies on both tetrahedral (Ga${(1)}$)
and octahedral (Ga$_{(2)}$) site are equally likely. Thus we need to address next why among the 
$V_\mathrm{Ga}$ models only vacancies on the Ga$_{(1)}$ sublattice would be seen in experiment. 
In $n$-type conducting $\beta$-Ga$_2$O$_3$
the $V_\mathrm{Ga}$ are expected to be all in the diamagnetic $q=-3$ charge state, with
all the O-dangling bond like states near the VBM completely filled, while
an EPR active $S=1/2$ state requires $q=-2$. 
The Fermi level position in equilibrium was estimated in Ref. \onlinecite{Deak17}.
After irradiation, equilibrium no longer applies and 
the Fermi level is expected to shift to a position  lower in the gap.
From the presence of the Fe$^{3+}$ EPR
and the relation of Fe$^{3+}$ with a Deep Level Transient Spectroscopy (DLTS)
level at 0.78 eV below the CBM\cite{Ingebrigtsen} we know it has to be at least
this deep.  Our hypothesis is now that the Fermi level position must be such
that only the M4 and M1  models are in the EPR active state, while the
other variants of the $V_\mathrm{Ga}$ are not. Since M4 is really the ground state configuration 
of M1, M4 would then be the preferred model. 
To establish this possibility we need to examine the transition levels. 
The calculated positions of the  transition levels for different relevant models
are shown in Fig. \ref{figtranslev} and given numerically in Table \ref{tabtranslev}. 
There is indeed a Fermi level range (as indicated by 'EPR' in Fig. \ref{figtranslev})
such that only the simple (M1) $V_\mathrm{Ga1}$ and the (b)-complex (M4) are in the EPR active state. 
We can see that both the $V_\mathrm{Ga1}$ and (b)-complex $V_\mathrm{Ga1}-\mathrm{Ga}_{ib}-V_\mathrm{Ga1}$
have  $2-/3-$ levels less deep below the conduction band minimum (CBM) at $-0.67$ eV and
$-0.74$ eV respectively. 
It is thus plausible that the $V_\mathrm{Ga2}$ and the (c)-complex (M5) are still in the 
$q=-3$ EPR inactive state if we pin the Fermi level in this range.
% UG: This info has been moved into caption of Fig. 9.
%We should note here that the transition level for the $V_\mathrm{Ga2}$ was calculated using 
%the titled spin model M3 for the $q=-2$  charge state which has the lowest energy in the hybrid 
%functional calculations for the 160 atom model. If the M2 model has higher energy, as is the
%case in hybrid functional, than its $2-/3-$ transition level would lie even deeper below the CBM.

The position of all transition levels is somewhat deeper in other calculations reported in the
literature.\cite{Varley11,Varley12,Kyrtsos17}
The reason for this discrepancy is a different dielectric constant (high-frequency instead of static) used in
De\'ak \etal's procedure\cite{Deak17} for correcting the image charge interactions.  
Irrespective of this choice, all calculations agree that the $V_\mathrm{Ga2}$ is deeper than the $V_\mathrm{Ga1}$, 
which is the crucial point needed here. We note that the estimated Fermi level position as indicated in Fig. 
\ref{figtranslev} should not prevent the possibility that the defect catches an additional hole leading to the 
corresponding $S=1$ state. Observation of such non-equilibrium states is not uncommon in EPR.

\subsection{Models for EPR2} 

\begin{figure*}
  \includegraphics[width=16cm]{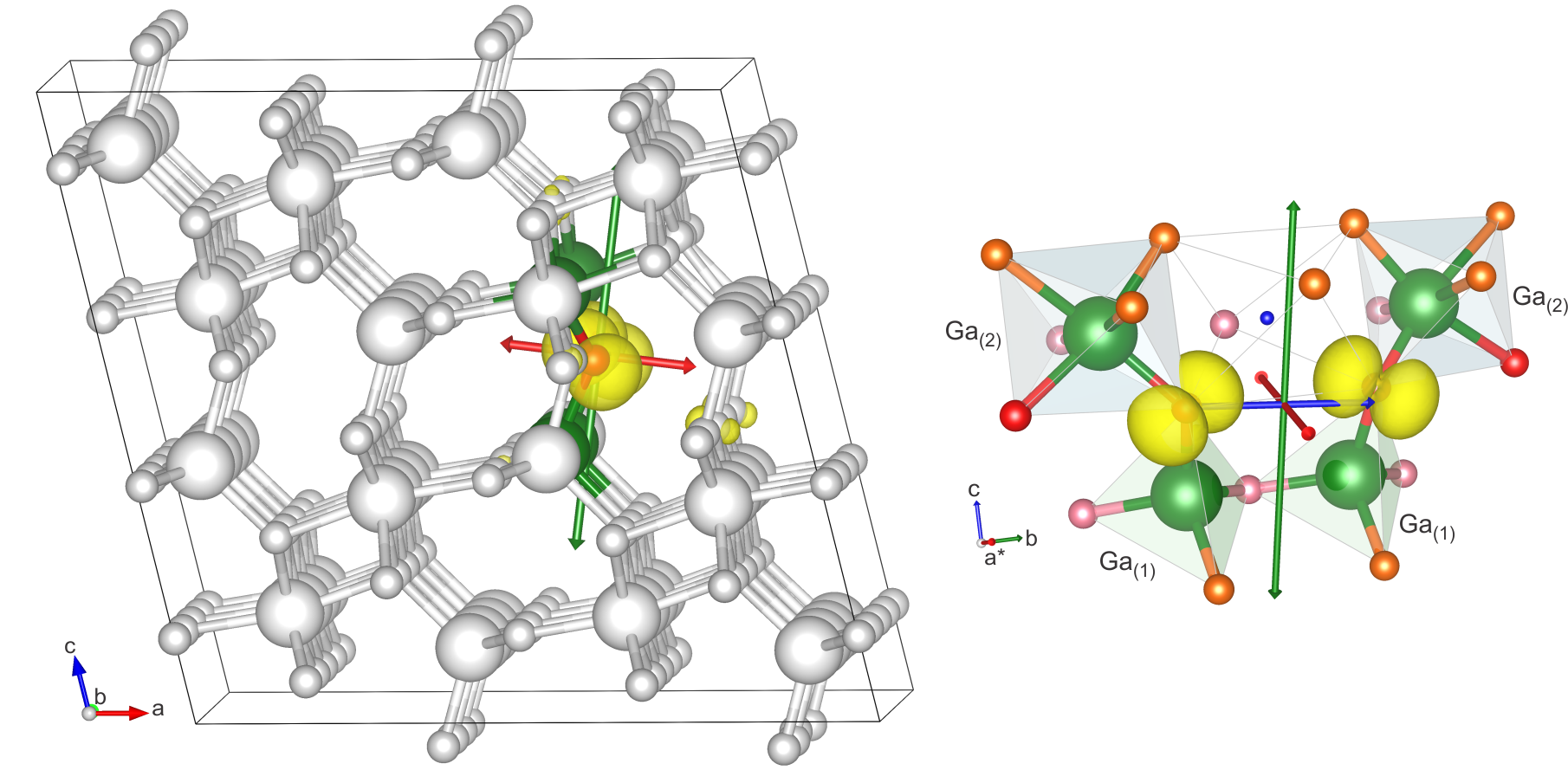} 
  \caption{Model M3 with tilted spins for $V_\mathrm{Ga2}$  structure,
    spin density, $g$-tensor and Ga atoms with strong HFI. Details as in
    Fig. \ref{figvga1}. Note that this is a symmetrized model in which spin appears to occur on both sides of the mirror plane of the defect. When relaxing this structure withouth symmetry constraint, the tilted spin occurs only on one side
    so that there is only hyperfine interaction with two Ga but the left
    and right variant would give the same ${\bf c}$-oriented major axis of the $g$-tensor. \label{figM3}}
\end{figure*}

Next, we consider the possible candidates for EPR2. Now that
we have established that the largest $g$-tensor direction  in these
defects tends to correspond to the pair of Ga connected to the spin carrying O atom,
it becomes clear that few candidates can lead to a $g$-tensor with maximum
along {\bf c}. Only the tilted spin model M3 (V$_\mathrm{Ga2}$) is a candidate among the
Ga-vacancy type defects. Its spin density and $g$-tensor and
HFI are shown in Fig. \ref{figM3}.
The reason is that in this case the directions of
the Ga pairs which is  in between {\bf b} and {\bf c} become averaged
over the two mirror-related pairs on either side of the mirror plane. Note that this
applies no matter whether one assumes a static Jahn-Teller effect or a dynamic
Jahn-Teller effect. In the former case, there will be some centers in the sample
where the spin is on the left and some where it is on the right and the
measurement would average over the two. 

The hyperfine splitting on the two Ga are not equal
for this case because they are inequivalent
structurally. For $U=4$ eV,  as given in Table \ref{tabcalceprvga}, they are
$-21$, $-16$ G. In hybrid functional they are $-33$ and $-22$ G.
The above results correspond to the symmetrized case where spin occurs
on both sides of the mirror plane and hyperfine
then would occur with 2 inequivalent pairs of Ga atoms.
Within PBE+$U$ with $U=4$ eV, however, the structure
actually breaks the symmetry and spin occurs only on one side of the
mirror plane. In that case, the difference between the two Ga hyperfine
becomes somewhat larger, $-25$ G, $-12.5$ G. This is related
to a larger difference in bond length of the spin-carrying O$_{(1)}$-Ga$_{(1)}$
and O$_{(1)}$-Ga$_{(2)}$ neighbors which are respectively closer and
farther from the mirror-plan.  
Interestingly, although the O-spin density becomes about a factor
two smaller when we split the spin equally over both sides, as is
to be expected, the Ga hyperfine changes much less. This illustrates
once more that the subtleties of how the wave function spreads to the
second neighbor Ga-$s$ and hence determines the SHF is not immediately
obvious from the overall localization behavior.  

The experiments were fit assuming two equivalent Ga and the characteristic fine
structure pattern was found to be quite sensitive to relaxing this assumption.
However, when we consider that both values are overestimated, their difference
may also become smaller in a more sophisticated description
including for example dynamic Jahn-Teller effects.
One would expect the DJT effect
to reduce the difference between the two Ga SHF and smear out the spectrum
while explaining nonetheless SHF with only 2 Ga atoms. 
The average value of the two inequivalent Ga is a bit
smaller than for the EPR1 model,
which is consistent with the experimental observation.
Upon close examination, one may notice that for EPR2
the details of the SHF structure are a bit less resolved than for EPR1 because
they show similar broadening but a smaller splitting. Thus a small
difference between the two Ga hyperfine cannot be excluded. Future experimental
work using higher microwave frequency might be able to reveal
the difference between the two Ga.

In terms of the $g$-tensor, this model has the largest
value along {\bf c} in agreement with experiment, but the next smaller value
is along ${\bf a}^*$ rather than {\bf b}. This small
deviation from experiment may be considered to fall within the error 
of the calculations. Thus, M3 appears to be a reasonable model for EPR2 but not perfect. 

Next, we discuss some of the alternative models based on self-trapped holes (STH) or 
oxygen interstitial (O$_i$).
First we consider the M7 self-trapped hole model with the hole trapped on O$_{(1)}$
which was previously proposed to be the origin of this spectrum.\cite{Kananensth}
Its $g$-tensor has  indeed principal value directions close to  {\bf c}, followed by {\bf b} 
and ${\bf a}^*$, however with values of $g_c\approx g_b\gg g_{a*}$. More importantly, there 
is strong hyperfine interaction with three Ga and the strongest component is actually on 
the third single Ga. 
The next model M8 with holes trapped on two adjacent O$_{(2)}$ has hyperfine with 6 Ga and 
not matching $g$-tensor and can, thus, easily be dismissed. A figure for this model can be 
found in Supplementary Information.\cite{supinfo}

\begin{figure*}
  \includegraphics[width=16cm]{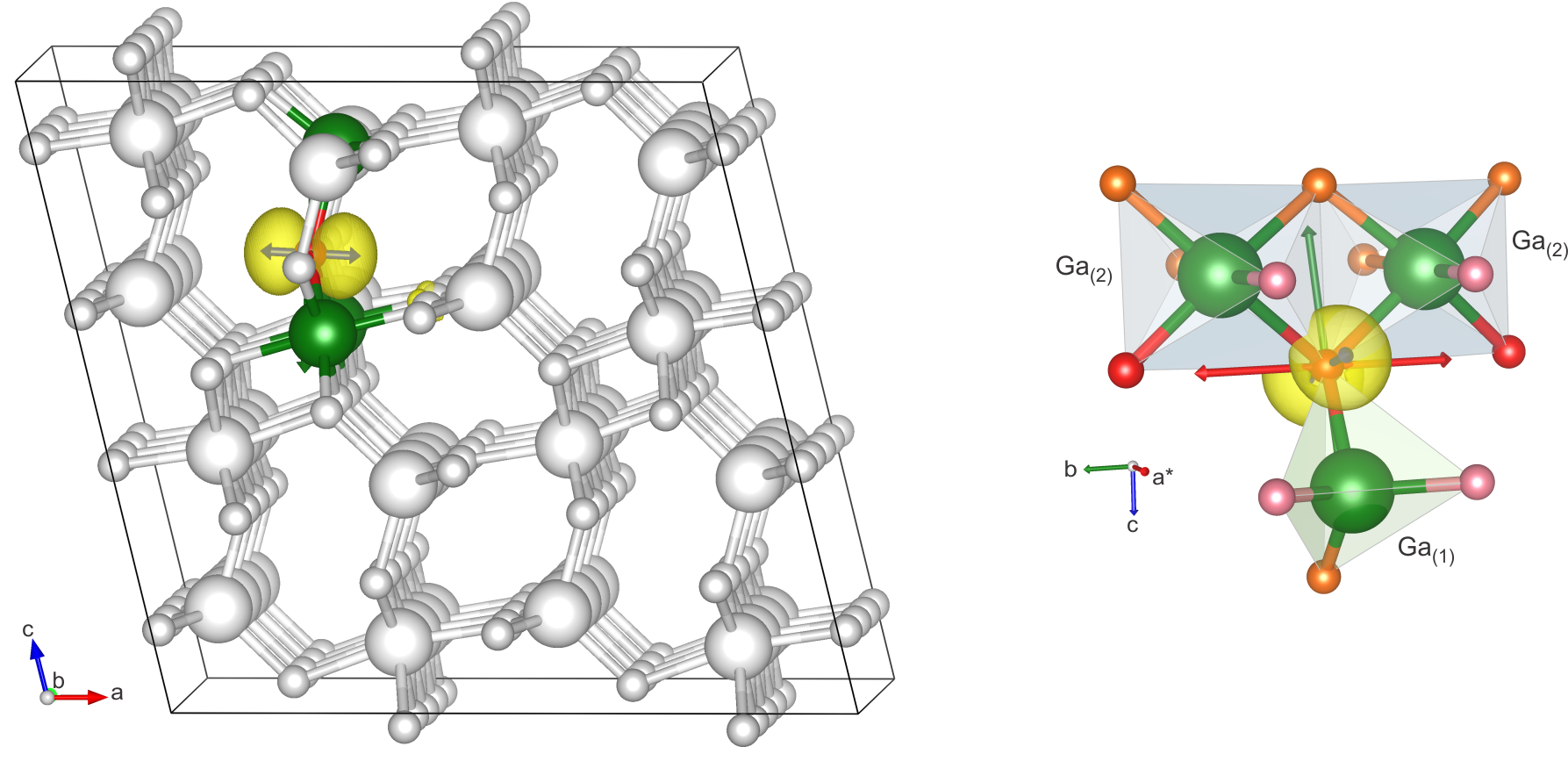} 
  \caption{Self-trapped polaron spin density on O$_{(1)}$ (Model M7)
    in yellow, $g$-tensor (double
    arrows) and Ga exhibiting strong HFI. Details as in Fig. \ref{figvga1}
    \label{figO1STH}}
\end{figure*}

\begin{figure*}
  \includegraphics[width=16cm]{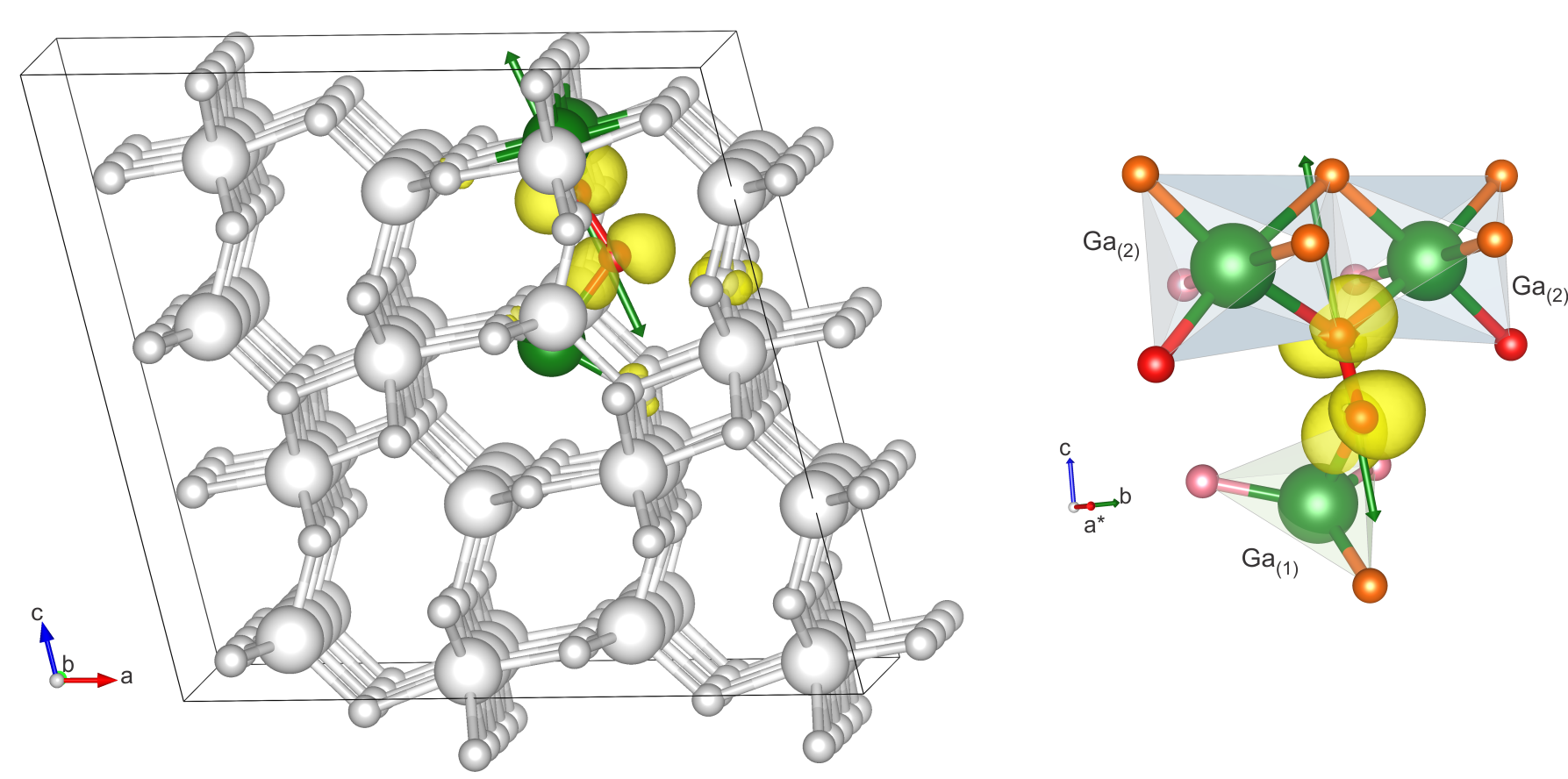} 
  \caption{Interstitial O$_i$ with O-O dumbbell oriented along {\bf c}
    (M9) structure,
 spin density in yellow, $g$-tensor (double
    arrows) and Ga exhibiting strong HFI. Details as in Fig. \ref{figvga1}
    \label{figM9}}
\end{figure*}

The O$_i$ model M9 on the other hand (shown in Fig. \ref{figM9})
has a $g$-tensor that matches the experiment
quite well in terms of the order of the  principal axes, $g_c\gg g_b\approx g_{a*}$ although $\Delta g_c$ is still a bit small.
However, the HFI is not in good agreement with experiment because  three Ga atoms have
significant HFI, an equivalent pair with value of $-8$ G  and a third one with a higher value of $-19$ G. 
Note that the values with the pair are significantly lower than for the $V_\mathrm{Ga}$. 
However, we cannot ignore the interaction with the third Ga because it is in fact the higher HFI.
As can be seen in Fig. \ref{figM9}, the strong HFI with the third Ga
results from the spin density to be in a $\pi$-like orbital
spread over the two O atoms in the dumbbell, as is to be expected.  
So, this model has the $g$-tensor which best matches the experiment but should be ruled
out because it would predict HFI with a prominent third Ga which would definitely change the
fine structure sufficiently to be detected experimentally.

As a further modification of this model, we removed the third Ga from M9 and thereby created 
model M10 ($\mathrm{O}_i-V_\mathrm{Ga1}$), shown in Supplementary Information.\cite{supinfo}
In that case, the spin becomes more pulled toward the O farther away from the Ga$_{(2)}$ pair, 
the $g$-tensor stays similar but the HFI on the pair of equivalent Ga becomes much too small.
These two O$_i$ models M9/M10 were relaxed only within DFT+$U$.
%than with hybrid functional, DFT$+U$ was found sufficiently accurate in other cases to
%predict the structure and these models are excluded for other reasons, so it did not
%seem warranted to recalculate them using the hybrid functional. 
%

Finally, referring back to Fig. \ref{figtranslev} we note that the self-trapped holes
(STH, models M7/M8) as well as the split-interstitial O$_i$ models (M9/M10) have much 
deeper transition levels to the EPR active positive charge states. While this does not 
completely exclude them under optical excitation, it makes them much less likely candidates. 

\subsection{$S=1$ spectra}
Several of the defects were calculated in the $q=-1$ state in order to
assign the $S=1$ spectra. It was found that for the
simple $V_\mathrm{Ga1}$ model, the second hole tends to localize on the
pair of O$_{(2)}$ rather than the O$_{(3)}$ on the mirror plane. This would break the symmetry
and not be consistent with an $S=1$ very closely related to the $S=1/2$ EPR1 spectrum. 
Similar considerations hold for $V_\mathrm{Ga2}$ when
starting from the symmetric solution with spin on O$_{(2)}$ on the mirror plane (M2).
On the other hand, the   $V_\mathrm{Ga1}-\mathrm{Ga}_{ib}-V_\mathrm{Ga1}$  (M4 model), which
we already consider as the most likely model for EPR1, is
an ideal candidate for the $S=1$ spectrum. First, the spins on the O$_{(1)}$ type atoms near each
$V_\mathrm{Ga1}$ in this model are far apart ($\sim$7.2 {\AA}) so that this would indeed correspond to two weakly
interacting spins.  Second, this $S=1$ $q=-1$ state was indeed found to have halved hyperfine interactions compared to the $S=1/2$ and a similar $g$-tensor with maximum along {\bf b} of $\Delta g_1=0.0316$, $\Delta g_2=0.014$ close to (about 17$^\circ$ from) the {\bf c}-axis and $\Delta g_3=0.007$ close to {\bf a*}. 
Third, this particular complex has lowest energy among 
all the Ga-vacancy related models in the $q=-1$ charge state (See Table \ref{tabtote}).

Next,  let us analyze the size of the the zero-field splitting.
The point dipole-dipole interaction between two $S=1/2$ spins is given by
\begin{equation}
  U_{12}({\bf r})=\frac{\mu_0}{4\pi}\frac{[{\bf m}_1\cdot{\bf m}_2-3({\bf m}_1\cdot\hat{\bf r})({\bf m}_2\cdot\hat{\bf r})]}{r^3}
\end{equation}
The splitting between the two  parts of the $S=1$ spectrum
is $2D$ and can be set equal to this energy difference if we assume
that the angular factors stay close to 1. 
Using $m_i=g_e\mu_B\sqrt{S/(S+1)}$ the $D$ value then corresponds to a
distance $d$ between the two $S=1/2$ of 5.4 \AA. This agrees with the
estimate by Kananen \etal\cite{Kananen17} The distance between the
two spin-carrying O atoms in the M4 model is 7.2 \AA. However,
the point dipole model is not sufficient to determine the distance accurately
to better than a few \AA\  because in a quantum mechanical calculations, one
would need to determine the expectation value of the dipole interaction for
the particular defect wave function of O-$p$ orbital.\cite{Riplinger}
Thus the observed value is consistent with the model within the error bars
of the point dipole model. The direction of the axial zero-field
splitting also agrees reasonably well with the vector linking
the two O atoms in the M4 model. 
We conclude that the occurrence of this $S=1$ spectrum and its characteristics
provides further strong confirmation
to our identification of EPR1 with the (b)-complex form of the $V_\mathrm{Ga1}$. 

As for the EPR2 related $S=1$ state, the M3 model is also a
plausible model to catch a second hole. 
We found the $S=1$ state of M3 with $q=-1$  to have a 
similar $g$-tensor: $\Delta g_1=0.037$, $\theta_1=90^\circ$, $\phi_1=56^\circ$, close to ${\bf c}$-axis,
$\Delta g_2=0.0245$, $\theta=0$ (${\bf b}$-axis), and $\Delta g_3=0.0154$, $\theta_3=90^\circ$, $\phi_3=-34^\circ$
(closer to ${\bf a}_*$) as the corresponding $S=1/2$ model. However,
there is somewhat larger interaction between the spins in this model because the O are only 3.5 {\AA} apart. This leads to a somewhat
larger difference between the nonequivalent Ga atoms
which now have hyperfine contact terms of
$-12$ G and $-27$ G instead of $-16$ G and $-21$ G in the $S=1/2$.  This is less clearly indicative of weakly coupled spins.
We note also that this newly found EPR2 $S=1$ spectrum has a more complex lineshape with a broad
line underlying the halved hyperfine splitting of the $S=1/2$ part of the spectrum.
This broadening could possibly be explained by the two sets of slightly
inequivalent pairs involved in this spectrum. The smaller distance is compatible with a larger zero field splitting
although the scaling with $d^{-3}$ would have suggested a larger difference.
Using the above approach, the $D=322$ MHz corresponds to 4.9 \AA\ instead
of the 3.5 \AA\ distance between the spins in the M3 model. 
We note once more that the point dipole-dipole interaction 
provides  only an approximate model for the zero-field splitting\cite{Riplinger} and also ignores the additional
exchange interaction between the spins, which might play a more significant role for these closer spins. The axis of the axial $D$ does  in this case not
correspond to the line joining the two O atoms but does lie in the
symmetry plane of the defect as expected for an axial $S=1$ center.

On the other hand,
the O$_i$ is very unlikely to have an associated $S=1$ $q=+2$ charge state.
The $q=+2$ state can safely be dismissed on the ground of total energy
calculations. This is another reason why this is a less attractive model for
EPR2. 

\subsection{Photoexcitation process}
Finally, the nature of the photoexcitation process and the meaning of the threshold value
are discussed. 
If we consider the EPR2 to be related to $V_\mathrm{Ga2}$ in the M3 model,
we may consider a migration path of Ga for converting the M4 model to
M3. Following Ref.~\onlinecite{Kyrtsos17} the migration barrier for
from Ga in the interstitial $i_b$ site to a Ga$_{(1)}$ site via a $q9$ jump is
1.0 eV in the $q=-2$ charge state.
(We follow here the notations of Kyrtsos \etal\cite{Kyrtsos17}.) We are then left with a single
$V_\mathrm{Ga1}$, which could migrate to one of the neighboring octahedral sites
by a $q6$, $q8$, $q10$ or $q4+q5$ paths.
These would add an additional 1.2-1.7 eV.
So, the total energy to transform the M4 to a $V_\mathrm{Ga2}$
path would amount to about $2.5\pm0.3$ eV. This agrees rather well
with the photothreshold. Also, within this process the defect configuration giving rise to EPR1
is expected to disappear at the same time as the configuraton giving EPR2 appears. 
In the above model,  the photoexcitation process is viewed as a direct transformation
of the defect from one form to the other related to $V_\mathrm{Ga}$ migration.

We could also view the photoexcitation process as transferring a hole
form an M4 model to an already existing nearby octahedral vacancy which is
still in the EPR inactive state. Or alternatively, the electron
is transferred from the octahedral to the tetrahedral vacancy  in its
complex M4 configuration. This would also simultaneously activate the octahedral
vacancy EPR signal and deactivate the M4 one as is observed to be the case. 

We should however keep in mind that  this EPR center
can also be created by X-ray absorption at low temperature starting
from the EPR1 containing sample.\cite{Kananensth}
So, the meaning of the photothreshold energy is not entirely clear.

\section{Conclusions}
In this paper, we provided experimental details of the EPR spectra previously
reported only briefly.\cite{Jurgen18} In particular,  we found that each spectrum
is accompanied by a corresponding $S=1$ spectrum. 
More importantly, we provide here the full detail of the calculations which leads to our
assignment of the EPR spectra to specific defect configurations.
When we allow for the still somewhat limited accuracy of PBE-calculated 
$g$-tensor values
and focus on the main qualitative features, such as the orientations of the principal axes of the
$g$-tensor relative to the crystal axes, and the number of Ga atoms on which hyperfine interaction
is found to be strong, a clear picture emerges.
The EPR1 spectrum most likely corresponds to the $V_\mathrm{Ga1}-\mathrm{Ga}_{ib}-V_\mathrm{Ga1}$ model,
which is in fact the ground state of $V_\mathrm{Ga1}$ in the EPR relevant $q=-2$ charge state.
The reason why other forms of the $V_\mathrm{Ga1}$ or the octahedral $V_\mathrm{Ga2}$ are not seen
in the experiment is that they are still in the EPR inactive $q=-3$ charge state even after
the Fermi level is somewhat lowered compared to the as grown crystal after the irradiation treatment. 
This explanation  is consistent with the calculated transition levels, which is indeed found to
be less deep below the conduction band for the  $V_\mathrm{Ga1}-\mathrm{Ga}_{ib}-V_\mathrm{Ga1}$ model
and with a plausible assignment of the Fermi level in these samples based on the observation of the
Fe$^{3+}$ EPR signal.

For EPR2, the EPR center found after photoexcitation,
the titled spin $V_\mathrm{Ga2}$
emerges as the most likely model.  While it agrees less closely with experimental values
than for EPR1, the alternatives considered here, such as the self-trapped hole or a split  interstitial 
oxygen could  all be ruled out based on various detailed arguments but mostly because they would clearly
show hyperfine interaction on three Ga atoms and are much less likely to be in an EPR active
charge state and cannot support a corresponding $S=1$ center. 

The hyperfine interaction for both models was found to be somewhat overestimated even when including
Hubbard-$U$ corrections to the GGA density functional or using a hybrid functional.
This indicates that calculating hyperfine interactions constitutes a very demanding test on the
accuracy of the  defect wave function. Another possible explanation for this discrepancy
could be the occurrence of a dynamic Jahn-Teller effect which may lead to a reduction of
the hyperfine interaction. This effect would apply to both of the here proposed models
which both correspond to a symmetry broken electronic state of the rather complex defect.

While the overestimate of the HFI certainly deserves further investigation as well as the
dynamical Jahn-Teller effect, it is not a crucial point in our assignment  of the
most plausible models to the spectra. This identification is mostly based on qualitative
features, such as the direction of the major axis of the $g$-tensor, the number of
Ga atoms which show hyperfine interaction, the likelihood of the different models
to be in the EPR active states based on their transition levels  and the capability of the
defect to capture a second hole leading to the corresponding $S=1$ spectra.
These aspects are robust and not sensitive to details of the calculation method.
Finally, we note that our identification of the EPR1 center with the $V_\mathrm{Ga1}-\mathrm{Ga}_{ib}-V_\mathrm{Ga1}$
  model provides an experimental confirmation of this interesting configuration of the Ga-vacancy
  in $\beta$-Ga$_2$O$_3$. Also, although we here focused on identifying the until now experimentally
  observed EPR spectra, the predictions for other defect models may become extremely useful in future
  experimental investigations of these materials, in particular those with modified Fermi level.  

\acknowledgements{The work at CWRU was supported by
  the National Science Foundation under grant No. DMR-1708593.
  The calculations were carried out at the Ohio Supercomputer Center under Project No. PDS0145. The work at BCCMS was supported by the DFG grant No. FR2833/63-1 and by the Supercomputer Center of Northern Germany (HLRN Grant No. hbc00027).
U.G. acknowledges support by the  Deutsche Forschungsgemeinschaft (DFG)
priority program SPP-1601.
We acknowledge the Helmholtz Zentrum Dresden (HZDR) in Rossendorf (Germany) for the proton and electron irradiation and the laboratory CEMHTI , CNRS, Orléans (France) for proton irradiation.}

\bibliography{ga2o3,dft,gipaw,hyperfine,ldau}

\begin{thebibliography}{51}%
\makeatletter
\providecommand \@ifxundefined [1]{%
 \@ifx{#1\undefined}
}%
\providecommand \@ifnum [1]{%
 \ifnum #1\expandafter \@firstoftwo
 \else \expandafter \@secondoftwo
 \fi
}%
\providecommand \@ifx [1]{%
 \ifx #1\expandafter \@firstoftwo
 \else \expandafter \@secondoftwo
 \fi
}%
\providecommand \natexlab [1]{#1}%
\providecommand \enquote  [1]{``#1''}%
\providecommand \bibnamefont  [1]{#1}%
\providecommand \bibfnamefont [1]{#1}%
\providecommand \citenamefont [1]{#1}%
\providecommand \href@noop [0]{\@secondoftwo}%
\providecommand \href [0]{\begingroup \@sanitize@url \@href}%
\providecommand \@href[1]{\@@startlink{#1}\@@href}%
\providecommand \@@href[1]{\endgroup#1\@@endlink}%
\providecommand \@sanitize@url [0]{\catcode `\\12\catcode `\$12\catcode
  `\&12\catcode `\#12\catcode `\^12\catcode `\_12\catcode `\%12\relax}%
\providecommand \@@startlink[1]{}%
\providecommand \@@endlink[0]{}%
\providecommand \url  [0]{\begingroup\@sanitize@url \@url }%
\providecommand \@url [1]{\endgroup\@href {#1}{\urlprefix }}%
\providecommand \urlprefix  [0]{URL }%
\providecommand \Eprint [0]{\href }%
\providecommand \doibase [0]{http://dx.doi.org/}%
\providecommand \selectlanguage [0]{\@gobble}%
\providecommand \bibinfo  [0]{\@secondoftwo}%
\providecommand \bibfield  [0]{\@secondoftwo}%
\providecommand \translation [1]{[#1]}%
\providecommand \BibitemOpen [0]{}%
\providecommand \bibitemStop [0]{}%
\providecommand \bibitemNoStop [0]{.\EOS\space}%
\providecommand \EOS [0]{\spacefactor3000\relax}%
\providecommand \BibitemShut  [1]{\csname bibitem#1\endcsname}%
\let\auto@bib@innerbib\@empty
%</preamble>
\bibitem [{\citenamefont {Sasaki}\ \emph {et~al.}(2013)\citenamefont {Sasaki},
  \citenamefont {Higashiwaki}, \citenamefont {Kuramata}, \citenamefont
  {Masui},\ and\ \citenamefont {Yamakoshi}}]{Sasaki13}%
  \BibitemOpen
  \bibfield  {author} {\bibinfo {author} {\bibfnamefont {K.}~\bibnamefont
  {Sasaki}}, \bibinfo {author} {\bibfnamefont {M.}~\bibnamefont {Higashiwaki}},
  \bibinfo {author} {\bibfnamefont {A.}~\bibnamefont {Kuramata}}, \bibinfo
  {author} {\bibfnamefont {T.}~\bibnamefont {Masui}}, \ and\ \bibinfo {author}
  {\bibfnamefont {S.}~\bibnamefont {Yamakoshi}},\ }\href {\doibase
  http://dx.doi.org/10.1016/j.jcrysgro.2013.02.015} {\bibfield  {journal}
  {\bibinfo  {journal} {J. Cryst. Growth}\ }\textbf {\bibinfo {volume} {378}},\
  \bibinfo {pages} {591 } (\bibinfo {year} {2013})},\ \bibinfo {note} {the 17th
  International Conference on Molecular Beam Epitaxy}\BibitemShut {NoStop}%
\bibitem [{\citenamefont {Matsumoto}\ \emph {et~al.}(1974)\citenamefont
  {Matsumoto}, \citenamefont {Aoki}, \citenamefont {Kinoshita},\ and\
  \citenamefont {Aono}}]{Matsumoto74}%
  \BibitemOpen
  \bibfield  {author} {\bibinfo {author} {\bibfnamefont {T.}~\bibnamefont
  {Matsumoto}}, \bibinfo {author} {\bibfnamefont {M.}~\bibnamefont {Aoki}},
  \bibinfo {author} {\bibfnamefont {A.}~\bibnamefont {Kinoshita}}, \ and\
  \bibinfo {author} {\bibfnamefont {T.}~\bibnamefont {Aono}},\ }\href
  {http://stacks.iop.org/1347-4065/13/i=10/a=1578} {\bibfield  {journal}
  {\bibinfo  {journal} {Jap. J. Appl. Phys.}\ }\textbf {\bibinfo {volume}
  {13}},\ \bibinfo {pages} {1578} (\bibinfo {year} {1974})}\BibitemShut
  {NoStop}%
\bibitem [{\citenamefont {Peelaers}\ and\ \citenamefont {Van~de
  Walle}(2015)}]{Peelaers15a}%
  \BibitemOpen
  \bibfield  {author} {\bibinfo {author} {\bibfnamefont {H.}~\bibnamefont
  {Peelaers}}\ and\ \bibinfo {author} {\bibfnamefont {C.~G.}\ \bibnamefont
  {Van~de Walle}},\ }\href {\doibase 10.1002/pssb.201451551} {\bibfield
  {journal} {\bibinfo  {journal} {Phys. Status Solidi (b)}\ }\textbf {\bibinfo
  {volume} {252}},\ \bibinfo {pages} {828} (\bibinfo {year}
  {2015})}\BibitemShut {NoStop}%
\bibitem [{\citenamefont {Furthm\"uller}\ and\ \citenamefont
  {Bechstedt}(2016)}]{Furthmuller16}%
  \BibitemOpen
  \bibfield  {author} {\bibinfo {author} {\bibfnamefont {J.}~\bibnamefont
  {Furthm\"uller}}\ and\ \bibinfo {author} {\bibfnamefont {F.}~\bibnamefont
  {Bechstedt}},\ }\href {\doibase 10.1103/PhysRevB.93.115204} {\bibfield
  {journal} {\bibinfo  {journal} {Phys. Rev. B}\ }\textbf {\bibinfo {volume}
  {93}},\ \bibinfo {pages} {115204} (\bibinfo {year} {2016})}\BibitemShut
  {NoStop}%
\bibitem [{\citenamefont {Mengle}\ \emph {et~al.}(2016)\citenamefont {Mengle},
  \citenamefont {Shi}, \citenamefont {Bayerl},\ and\ \citenamefont
  {Kioupakis}}]{Mengle16}%
  \BibitemOpen
  \bibfield  {author} {\bibinfo {author} {\bibfnamefont {K.~A.}\ \bibnamefont
  {Mengle}}, \bibinfo {author} {\bibfnamefont {G.}~\bibnamefont {Shi}},
  \bibinfo {author} {\bibfnamefont {D.}~\bibnamefont {Bayerl}}, \ and\ \bibinfo
  {author} {\bibfnamefont {E.}~\bibnamefont {Kioupakis}},\ }\href {\doibase
  10.1063/1.4968822} {\bibfield  {journal} {\bibinfo  {journal} {Appl. Phys.
  Lett.}\ }\textbf {\bibinfo {volume} {109}},\ \bibinfo {pages} {212104}
  (\bibinfo {year} {2016})},\ \Eprint
  {http://arxiv.org/abs/http://dx.doi.org/10.1063/1.4968822}
  {http://dx.doi.org/10.1063/1.4968822} \BibitemShut {NoStop}%
\bibitem [{\citenamefont {Ratnaparkhe}\ and\ \citenamefont
  {Lambrecht}(2017)}]{Ratnaparkhe17}%
  \BibitemOpen
  \bibfield  {author} {\bibinfo {author} {\bibfnamefont {A.}~\bibnamefont
  {Ratnaparkhe}}\ and\ \bibinfo {author} {\bibfnamefont {W.~R.~L.}\
  \bibnamefont {Lambrecht}},\ }\href {\doibase 10.1063/1.4978668} {\bibfield
  {journal} {\bibinfo  {journal} {Applied Physics Letters}\ }\textbf {\bibinfo
  {volume} {110}},\ \bibinfo {pages} {132103} (\bibinfo {year} {2017})},\
  \Eprint {http://arxiv.org/abs/http://dx.doi.org/10.1063/1.4978668}
  {http://dx.doi.org/10.1063/1.4978668} \BibitemShut {NoStop}%
\bibitem [{\citenamefont {Green}\ \emph {et~al.}(2016)\citenamefont {Green},
  \citenamefont {Chabak}, \citenamefont {Heller}, \citenamefont {Fitch},
  \citenamefont {Baldini}, \citenamefont {Fiedler}, \citenamefont {Irmscher},
  \citenamefont {Wagner}, \citenamefont {Galazka}, \citenamefont {Tetlak},
  \citenamefont {Crespo}, \citenamefont {Leedy},\ and\ \citenamefont
  {Jessen}}]{Green16}%
  \BibitemOpen
  \bibfield  {author} {\bibinfo {author} {\bibfnamefont {A.~J.}\ \bibnamefont
  {Green}}, \bibinfo {author} {\bibfnamefont {K.~D.}\ \bibnamefont {Chabak}},
  \bibinfo {author} {\bibfnamefont {E.~R.}\ \bibnamefont {Heller}}, \bibinfo
  {author} {\bibfnamefont {R.~C.}\ \bibnamefont {Fitch}}, \bibinfo {author}
  {\bibfnamefont {M.}~\bibnamefont {Baldini}}, \bibinfo {author} {\bibfnamefont
  {A.}~\bibnamefont {Fiedler}}, \bibinfo {author} {\bibfnamefont
  {K.}~\bibnamefont {Irmscher}}, \bibinfo {author} {\bibfnamefont
  {G.}~\bibnamefont {Wagner}}, \bibinfo {author} {\bibfnamefont
  {Z.}~\bibnamefont {Galazka}}, \bibinfo {author} {\bibfnamefont {S.~E.}\
  \bibnamefont {Tetlak}}, \bibinfo {author} {\bibfnamefont {A.}~\bibnamefont
  {Crespo}}, \bibinfo {author} {\bibfnamefont {K.}~\bibnamefont {Leedy}}, \
  and\ \bibinfo {author} {\bibfnamefont {G.~H.}\ \bibnamefont {Jessen}},\
  }\href {\doibase 10.1109/LED.2016.2568139} {\bibfield  {journal} {\bibinfo
  {journal} {IEEE Electron Device Letters}\ }\textbf {\bibinfo {volume} {37}},\
  \bibinfo {pages} {902} (\bibinfo {year} {2016})}\BibitemShut {NoStop}%
\bibitem [{\citenamefont {Baliga}(1989)}]{Baliga}%
  \BibitemOpen
  \bibfield  {author} {\bibinfo {author} {\bibfnamefont {B.~J.}\ \bibnamefont
  {Baliga}},\ }\href {\doibase 10.1109/55.43098} {\bibfield  {journal}
  {\bibinfo  {journal} {IEEE Electron Device Letters}\ }\textbf {\bibinfo
  {volume} {10}},\ \bibinfo {pages} {455} (\bibinfo {year} {1989})}\BibitemShut
  {NoStop}%
\bibitem [{\citenamefont {Peelaers}\ \emph {et~al.}(2015)\citenamefont
  {Peelaers}, \citenamefont {Steiauf}, \citenamefont {Varley}, \citenamefont
  {Janotti},\ and\ \citenamefont {Van~de Walle}}]{Peelaers15}%
  \BibitemOpen
  \bibfield  {author} {\bibinfo {author} {\bibfnamefont {H.}~\bibnamefont
  {Peelaers}}, \bibinfo {author} {\bibfnamefont {D.}~\bibnamefont {Steiauf}},
  \bibinfo {author} {\bibfnamefont {J.~B.}\ \bibnamefont {Varley}}, \bibinfo
  {author} {\bibfnamefont {A.}~\bibnamefont {Janotti}}, \ and\ \bibinfo
  {author} {\bibfnamefont {C.~G.}\ \bibnamefont {Van~de Walle}},\ }\href
  {\doibase 10.1103/PhysRevB.92.085206} {\bibfield  {journal} {\bibinfo
  {journal} {Phys. Rev. B}\ }\textbf {\bibinfo {volume} {92}},\ \bibinfo
  {pages} {085206} (\bibinfo {year} {2015})}\BibitemShut {NoStop}%
\bibitem [{\citenamefont {Varley}\ \emph {et~al.}(2011)\citenamefont {Varley},
  \citenamefont {Peelaers}, \citenamefont {Janotti},\ and\ \citenamefont
  {Van~de Walle}}]{Varley11}%
  \BibitemOpen
  \bibfield  {author} {\bibinfo {author} {\bibfnamefont {J.~B.}\ \bibnamefont
  {Varley}}, \bibinfo {author} {\bibfnamefont {H.}~\bibnamefont {Peelaers}},
  \bibinfo {author} {\bibfnamefont {A.}~\bibnamefont {Janotti}}, \ and\
  \bibinfo {author} {\bibfnamefont {C.~G.}\ \bibnamefont {Van~de Walle}},\
  }\href {\doibase 10.1088/0953-8984/23/33/334212} {\bibfield  {journal}
  {\bibinfo  {journal} {J. Phys. Condens. Matter}\ }\textbf {\bibinfo {volume}
  {23}},\ \bibinfo {pages} {334212} (\bibinfo {year} {2011})}\BibitemShut
  {NoStop}%
\bibitem [{\citenamefont {Varley}\ \emph {et~al.}(2012)\citenamefont {Varley},
  \citenamefont {Janotti}, \citenamefont {Franchini},\ and\ \citenamefont
  {Van~de Walle}}]{Varley12}%
  \BibitemOpen
  \bibfield  {author} {\bibinfo {author} {\bibfnamefont {J.~B.}\ \bibnamefont
  {Varley}}, \bibinfo {author} {\bibfnamefont {A.}~\bibnamefont {Janotti}},
  \bibinfo {author} {\bibfnamefont {C.}~\bibnamefont {Franchini}}, \ and\
  \bibinfo {author} {\bibfnamefont {C.~G.}\ \bibnamefont {Van~de Walle}},\
  }\href {\doibase 10.1103/PhysRevB.85.081109} {\bibfield  {journal} {\bibinfo
  {journal} {Phys. Rev. B}\ }\textbf {\bibinfo {volume} {85}},\ \bibinfo
  {pages} {081109} (\bibinfo {year} {2012})}\BibitemShut {NoStop}%
\bibitem [{\citenamefont {Harwig}\ and\ \citenamefont
  {Kellendonk}(1978)}]{Harwig78}%
  \BibitemOpen
  \bibfield  {author} {\bibinfo {author} {\bibfnamefont {T.}~\bibnamefont
  {Harwig}}\ and\ \bibinfo {author} {\bibfnamefont {F.}~\bibnamefont
  {Kellendonk}},\ }\href {\doibase
  http://dx.doi.org/10.1016/0022-4596(78)90017-8} {\bibfield  {journal}
  {\bibinfo  {journal} {J. Solid State Chem.}\ }\textbf {\bibinfo {volume}
  {24}},\ \bibinfo {pages} {255 } (\bibinfo {year} {1978})}\BibitemShut
  {NoStop}%
\bibitem [{\citenamefont {Zacherle}\ \emph {et~al.}(2013)\citenamefont
  {Zacherle}, \citenamefont {Schmidt},\ and\ \citenamefont
  {Martin}}]{Zacherle13}%
  \BibitemOpen
  \bibfield  {author} {\bibinfo {author} {\bibfnamefont {T.}~\bibnamefont
  {Zacherle}}, \bibinfo {author} {\bibfnamefont {P.~C.}\ \bibnamefont
  {Schmidt}}, \ and\ \bibinfo {author} {\bibfnamefont {M.}~\bibnamefont
  {Martin}},\ }\href {\doibase 10.1103/PhysRevB.87.235206} {\bibfield
  {journal} {\bibinfo  {journal} {Phys. Rev. B}\ }\textbf {\bibinfo {volume}
  {87}},\ \bibinfo {pages} {235206} (\bibinfo {year} {2013})}\BibitemShut
  {NoStop}%
\bibitem [{\citenamefont {Peelaers}\ and\ \citenamefont {Van~de
  Walle}(2016)}]{Peelaers16}%
  \BibitemOpen
  \bibfield  {author} {\bibinfo {author} {\bibfnamefont {H.}~\bibnamefont
  {Peelaers}}\ and\ \bibinfo {author} {\bibfnamefont {C.~G.}\ \bibnamefont
  {Van~de Walle}},\ }\href {\doibase 10.1103/PhysRevB.94.195203} {\bibfield
  {journal} {\bibinfo  {journal} {Phys. Rev. B}\ }\textbf {\bibinfo {volume}
  {94}},\ \bibinfo {pages} {195203} (\bibinfo {year} {2016})}\BibitemShut
  {NoStop}%
\bibitem [{\citenamefont {De\'ak}\ \emph {et~al.}(2017)\citenamefont {De\'ak},
  \citenamefont {Duy~Ho}, \citenamefont {Seemann}, \citenamefont {Aradi},
  \citenamefont {Lorke},\ and\ \citenamefont {Frauenheim}}]{Deak17}%
  \BibitemOpen
  \bibfield  {author} {\bibinfo {author} {\bibfnamefont {P.}~\bibnamefont
  {De\'ak}}, \bibinfo {author} {\bibfnamefont {Q.}~\bibnamefont {Duy~Ho}},
  \bibinfo {author} {\bibfnamefont {F.}~\bibnamefont {Seemann}}, \bibinfo
  {author} {\bibfnamefont {B.}~\bibnamefont {Aradi}}, \bibinfo {author}
  {\bibfnamefont {M.}~\bibnamefont {Lorke}}, \ and\ \bibinfo {author}
  {\bibfnamefont {T.}~\bibnamefont {Frauenheim}},\ }\href {\doibase
  10.1103/PhysRevB.95.075208} {\bibfield  {journal} {\bibinfo  {journal} {Phys.
  Rev. B}\ }\textbf {\bibinfo {volume} {95}},\ \bibinfo {pages} {075208}
  (\bibinfo {year} {2017})}\BibitemShut {NoStop}%
\bibitem [{\citenamefont {Ingebrigtsen}\ \emph {et~al.}(2018)\citenamefont
  {Ingebrigtsen}, \citenamefont {Varley}, \citenamefont {Kuznetsov},
  \citenamefont {Svensson}, \citenamefont {Alfieri}, \citenamefont {Mihaila},
  \citenamefont {Badst\"ubner},\ and\ \citenamefont {Vines}}]{Ingebrigtsen}%
  \BibitemOpen
  \bibfield  {author} {\bibinfo {author} {\bibfnamefont {M.~E.}\ \bibnamefont
  {Ingebrigtsen}}, \bibinfo {author} {\bibfnamefont {J.~B.}\ \bibnamefont
  {Varley}}, \bibinfo {author} {\bibfnamefont {A.~Y.}\ \bibnamefont
  {Kuznetsov}}, \bibinfo {author} {\bibfnamefont {B.~G.}\ \bibnamefont
  {Svensson}}, \bibinfo {author} {\bibfnamefont {G.}~\bibnamefont {Alfieri}},
  \bibinfo {author} {\bibfnamefont {A.}~\bibnamefont {Mihaila}}, \bibinfo
  {author} {\bibfnamefont {U.}~\bibnamefont {Badst\"ubner}}, \ and\ \bibinfo
  {author} {\bibfnamefont {L.}~\bibnamefont {Vines}},\ }\href {\doibase
  10.1063/1.5020134} {\bibfield  {journal} {\bibinfo  {journal} {Applied
  Physics Letters}\ }\textbf {\bibinfo {volume} {112}},\ \bibinfo {pages}
  {042104} (\bibinfo {year} {2018})},\ \Eprint
  {http://arxiv.org/abs/https://doi.org/10.1063/1.5020134}
  {https://doi.org/10.1063/1.5020134} \BibitemShut {NoStop}%
\bibitem [{\citenamefont {Irmscher}\ \emph {et~al.}(2011)\citenamefont
  {Irmscher}, \citenamefont {Galazka}, \citenamefont {Pietsch}, \citenamefont
  {Uecker},\ and\ \citenamefont {Fornari}}]{Irmscher11}%
  \BibitemOpen
  \bibfield  {author} {\bibinfo {author} {\bibfnamefont {K.}~\bibnamefont
  {Irmscher}}, \bibinfo {author} {\bibfnamefont {Z.}~\bibnamefont {Galazka}},
  \bibinfo {author} {\bibfnamefont {M.}~\bibnamefont {Pietsch}}, \bibinfo
  {author} {\bibfnamefont {R.}~\bibnamefont {Uecker}}, \ and\ \bibinfo {author}
  {\bibfnamefont {R.}~\bibnamefont {Fornari}},\ }\href {\doibase
  10.1063/1.3642962} {\bibfield  {journal} {\bibinfo  {journal} {Journal of
  Applied Physics}\ }\textbf {\bibinfo {volume} {110}},\ \bibinfo {pages}
  {063720} (\bibinfo {year} {2011})},\ \Eprint
  {http://arxiv.org/abs/https://doi.org/10.1063/1.3642962}
  {https://doi.org/10.1063/1.3642962} \BibitemShut {NoStop}%
\bibitem [{\citenamefont {Zhang}\ \emph {et~al.}(2016)\citenamefont {Zhang},
  \citenamefont {Farzana}, \citenamefont {Arehart},\ and\ \citenamefont
  {Ringel}}]{Zhang16}%
  \BibitemOpen
  \bibfield  {author} {\bibinfo {author} {\bibfnamefont {Z.}~\bibnamefont
  {Zhang}}, \bibinfo {author} {\bibfnamefont {E.}~\bibnamefont {Farzana}},
  \bibinfo {author} {\bibfnamefont {A.~R.}\ \bibnamefont {Arehart}}, \ and\
  \bibinfo {author} {\bibfnamefont {S.~A.}\ \bibnamefont {Ringel}},\ }\href
  {\doibase 10.1063/1.4941429} {\bibfield  {journal} {\bibinfo  {journal}
  {Applied Physics Letters}\ }\textbf {\bibinfo {volume} {108}},\ \bibinfo
  {pages} {052105} (\bibinfo {year} {2016})},\ \Eprint
  {http://arxiv.org/abs/https://doi.org/10.1063/1.4941429}
  {https://doi.org/10.1063/1.4941429} \BibitemShut {NoStop}%
\bibitem [{\citenamefont {Ingebrigtsen}\ \emph {et~al.}(2019)\citenamefont
  {Ingebrigtsen}, \citenamefont {Kuznetsov}, \citenamefont {Svensson},
  \citenamefont {Alfieri}, \citenamefont {Mihaila}, \citenamefont
  {Badstübner}, \citenamefont {Perron}, \citenamefont {Vines},\ and\
  \citenamefont {Varley}}]{Ingebrigtsen19}%
  \BibitemOpen
  \bibfield  {author} {\bibinfo {author} {\bibfnamefont {M.~E.}\ \bibnamefont
  {Ingebrigtsen}}, \bibinfo {author} {\bibfnamefont {A.~Y.}\ \bibnamefont
  {Kuznetsov}}, \bibinfo {author} {\bibfnamefont {B.~G.}\ \bibnamefont
  {Svensson}}, \bibinfo {author} {\bibfnamefont {G.}~\bibnamefont {Alfieri}},
  \bibinfo {author} {\bibfnamefont {A.}~\bibnamefont {Mihaila}}, \bibinfo
  {author} {\bibfnamefont {U.}~\bibnamefont {Badstübner}}, \bibinfo {author}
  {\bibfnamefont {A.}~\bibnamefont {Perron}}, \bibinfo {author} {\bibfnamefont
  {L.}~\bibnamefont {Vines}}, \ and\ \bibinfo {author} {\bibfnamefont {J.~B.}\
  \bibnamefont {Varley}},\ }\href {\doibase 10.1063/1.5054826} {\bibfield
  {journal} {\bibinfo  {journal} {APL Materials}\ }\textbf {\bibinfo {volume}
  {7}},\ \bibinfo {pages} {022510} (\bibinfo {year} {2019})},\ \Eprint
  {http://arxiv.org/abs/https://doi.org/10.1063/1.5054826}
  {https://doi.org/10.1063/1.5054826} \BibitemShut {NoStop}%
\bibitem [{\citenamefont {Gao}\ \emph {et~al.}(2018)\citenamefont {Gao},
  \citenamefont {Muralidharan}, \citenamefont {Pronin}, \citenamefont {Karim},
  \citenamefont {White}, \citenamefont {Asel}, \citenamefont {Foster},
  \citenamefont {Krishnamoorthy}, \citenamefont {Rajan}, \citenamefont {Cao},
  \citenamefont {Higashiwaki}, \citenamefont {von Wenckstern}, \citenamefont
  {Grundmann}, \citenamefont {Zhao}, \citenamefont {Look},\ and\ \citenamefont
  {Brillson}}]{Gao18}%
  \BibitemOpen
  \bibfield  {author} {\bibinfo {author} {\bibfnamefont {H.}~\bibnamefont
  {Gao}}, \bibinfo {author} {\bibfnamefont {S.}~\bibnamefont {Muralidharan}},
  \bibinfo {author} {\bibfnamefont {N.}~\bibnamefont {Pronin}}, \bibinfo
  {author} {\bibfnamefont {M.~R.}\ \bibnamefont {Karim}}, \bibinfo {author}
  {\bibfnamefont {S.~M.}\ \bibnamefont {White}}, \bibinfo {author}
  {\bibfnamefont {T.}~\bibnamefont {Asel}}, \bibinfo {author} {\bibfnamefont
  {G.}~\bibnamefont {Foster}}, \bibinfo {author} {\bibfnamefont
  {S.}~\bibnamefont {Krishnamoorthy}}, \bibinfo {author} {\bibfnamefont
  {S.}~\bibnamefont {Rajan}}, \bibinfo {author} {\bibfnamefont {L.~R.}\
  \bibnamefont {Cao}}, \bibinfo {author} {\bibfnamefont {M.}~\bibnamefont
  {Higashiwaki}}, \bibinfo {author} {\bibfnamefont {H.}~\bibnamefont {von
  Wenckstern}}, \bibinfo {author} {\bibfnamefont {M.}~\bibnamefont
  {Grundmann}}, \bibinfo {author} {\bibfnamefont {H.}~\bibnamefont {Zhao}},
  \bibinfo {author} {\bibfnamefont {D.~C.}\ \bibnamefont {Look}}, \ and\
  \bibinfo {author} {\bibfnamefont {L.~J.}\ \bibnamefont {Brillson}},\ }\href
  {\doibase 10.1063/1.5026770} {\bibfield  {journal} {\bibinfo  {journal}
  {Applied Physics Letters}\ }\textbf {\bibinfo {volume} {112}},\ \bibinfo
  {pages} {242102} (\bibinfo {year} {2018})},\ \Eprint
  {http://arxiv.org/abs/https://doi.org/10.1063/1.5026770}
  {https://doi.org/10.1063/1.5026770} \BibitemShut {NoStop}%
\bibitem [{\citenamefont {Islam}\ \emph {et~al.}(2019)\citenamefont {Islam},
  \citenamefont {Rana}, \citenamefont {Hernandez}, \citenamefont {Haseman},\
  and\ \citenamefont {Selim}}]{Islam19}%
  \BibitemOpen
  \bibfield  {author} {\bibinfo {author} {\bibfnamefont {M.~M.}\ \bibnamefont
  {Islam}}, \bibinfo {author} {\bibfnamefont {D.}~\bibnamefont {Rana}},
  \bibinfo {author} {\bibfnamefont {A.}~\bibnamefont {Hernandez}}, \bibinfo
  {author} {\bibfnamefont {M.}~\bibnamefont {Haseman}}, \ and\ \bibinfo
  {author} {\bibfnamefont {F.~A.}\ \bibnamefont {Selim}},\ }\href {\doibase
  10.1063/1.5066424} {\bibfield  {journal} {\bibinfo  {journal} {Journal of
  Applied Physics}\ }\textbf {\bibinfo {volume} {125}},\ \bibinfo {pages}
  {055701} (\bibinfo {year} {2019})},\ \Eprint
  {http://arxiv.org/abs/https://doi.org/10.1063/1.5066424}
  {https://doi.org/10.1063/1.5066424} \BibitemShut {NoStop}%
\bibitem [{\citenamefont {Kananen}\ \emph
  {et~al.}(2017{\natexlab{a}})\citenamefont {Kananen}, \citenamefont
  {Halliburton}, \citenamefont {Stevens}, \citenamefont {Foundos},\ and\
  \citenamefont {Giles}}]{Kananen17}%
  \BibitemOpen
  \bibfield  {author} {\bibinfo {author} {\bibfnamefont {B.~E.}\ \bibnamefont
  {Kananen}}, \bibinfo {author} {\bibfnamefont {L.~E.}\ \bibnamefont
  {Halliburton}}, \bibinfo {author} {\bibfnamefont {K.~T.}\ \bibnamefont
  {Stevens}}, \bibinfo {author} {\bibfnamefont {G.~K.}\ \bibnamefont
  {Foundos}}, \ and\ \bibinfo {author} {\bibfnamefont {N.~C.}\ \bibnamefont
  {Giles}},\ }\href {\doibase 10.1063/1.4983814} {\bibfield  {journal}
  {\bibinfo  {journal} {Applied Physics Letters}\ }\textbf {\bibinfo {volume}
  {110}},\ \bibinfo {pages} {202104} (\bibinfo {year} {2017}{\natexlab{a}})},\
  \Eprint {http://arxiv.org/abs/http://dx.doi.org/10.1063/1.4983814}
  {http://dx.doi.org/10.1063/1.4983814} \BibitemShut {NoStop}%
\bibitem [{\citenamefont {Kananen}\ \emph
  {et~al.}(2017{\natexlab{b}})\citenamefont {Kananen}, \citenamefont {Giles},
  \citenamefont {Halliburton}, \citenamefont {Foundos}, \citenamefont {Chang},\
  and\ \citenamefont {Stevens}}]{Kananensth}%
  \BibitemOpen
  \bibfield  {author} {\bibinfo {author} {\bibfnamefont {B.~E.}\ \bibnamefont
  {Kananen}}, \bibinfo {author} {\bibfnamefont {N.~C.}\ \bibnamefont {Giles}},
  \bibinfo {author} {\bibfnamefont {L.~E.}\ \bibnamefont {Halliburton}},
  \bibinfo {author} {\bibfnamefont {G.~K.}\ \bibnamefont {Foundos}}, \bibinfo
  {author} {\bibfnamefont {K.~B.}\ \bibnamefont {Chang}}, \ and\ \bibinfo
  {author} {\bibfnamefont {K.~T.}\ \bibnamefont {Stevens}},\ }\href {\doibase
  10.1063/1.5007095} {\bibfield  {journal} {\bibinfo  {journal} {Journal of
  Applied Physics}\ }\textbf {\bibinfo {volume} {122}},\ \bibinfo {pages}
  {215703} (\bibinfo {year} {2017}{\natexlab{b}})},\ \Eprint
  {http://arxiv.org/abs/https://doi.org/10.1063/1.5007095}
  {https://doi.org/10.1063/1.5007095} \BibitemShut {NoStop}%
\bibitem [{\citenamefont {von Bardeleben}\ \emph {et~al.}(2019)\citenamefont
  {von Bardeleben}, \citenamefont {Zhou}, \citenamefont {Gerstmann},
  \citenamefont {Skachkov}, \citenamefont {Lambrecht}, \citenamefont {Ho},\
  and\ \citenamefont {De\'ak}}]{Jurgen18}%
  \BibitemOpen
  \bibfield  {author} {\bibinfo {author} {\bibfnamefont {H.~J.}\ \bibnamefont
  {von Bardeleben}}, \bibinfo {author} {\bibfnamefont {S.}~\bibnamefont
  {Zhou}}, \bibinfo {author} {\bibfnamefont {U.}~\bibnamefont {Gerstmann}},
  \bibinfo {author} {\bibfnamefont {D.}~\bibnamefont {Skachkov}}, \bibinfo
  {author} {\bibfnamefont {W.~R.~L.}\ \bibnamefont {Lambrecht}}, \bibinfo
  {author} {\bibfnamefont {Q.~D.}\ \bibnamefont {Ho}}, \ and\ \bibinfo {author}
  {\bibfnamefont {P.}~\bibnamefont {De\'ak}},\ }\href {\doibase
  10.1063/1.5053158} {\bibfield  {journal} {\bibinfo  {journal} {APL
  Materials}\ }\textbf {\bibinfo {volume} {7}},\ \bibinfo {pages} {022521}
  (\bibinfo {year} {2019})},\ \Eprint
  {http://arxiv.org/abs/https://doi.org/10.1063/1.5053158}
  {https://doi.org/10.1063/1.5053158} \BibitemShut {NoStop}%
\bibitem [{\citenamefont {Geller}(1960)}]{Geller60}%
  \BibitemOpen
  \bibfield  {author} {\bibinfo {author} {\bibfnamefont {S.}~\bibnamefont
  {Geller}},\ }\href {\doibase 10.1063/1.1731237} {\bibfield  {journal}
  {\bibinfo  {journal} {J. Chem. Phys.}\ }\textbf {\bibinfo {volume} {33}},\
  \bibinfo {pages} {676} (\bibinfo {year} {1960})},\ \Eprint
  {http://arxiv.org/abs/http://dx.doi.org/10.1063/1.1731237}
  {http://dx.doi.org/10.1063/1.1731237} \BibitemShut {NoStop}%
\bibitem [{\citenamefont {Perdew}\ \emph
  {et~al.}(1996{\natexlab{a}})\citenamefont {Perdew}, \citenamefont
  {Ernzerhof},\ and\ \citenamefont {Burke}}]{PBEh}%
  \BibitemOpen
  \bibfield  {author} {\bibinfo {author} {\bibfnamefont {J.~P.}\ \bibnamefont
  {Perdew}}, \bibinfo {author} {\bibfnamefont {M.}~\bibnamefont {Ernzerhof}}, \
  and\ \bibinfo {author} {\bibfnamefont {K.}~\bibnamefont {Burke}},\ }\href
  {\doibase 10.1063/1.472933} {\bibfield  {journal} {\bibinfo  {journal} {J.
  Chem. Phys.}\ }\textbf {\bibinfo {volume} {105}},\ \bibinfo {pages} {9982}
  (\bibinfo {year} {1996}{\natexlab{a}})}\BibitemShut {NoStop}%
\bibitem [{\citenamefont {Heyd}\ \emph {et~al.}(2003)\citenamefont {Heyd},
  \citenamefont {Scuseria},\ and\ \citenamefont {Ernzerhof}}]{HSE03}%
  \BibitemOpen
  \bibfield  {author} {\bibinfo {author} {\bibfnamefont {J.}~\bibnamefont
  {Heyd}}, \bibinfo {author} {\bibfnamefont {G.~E.}\ \bibnamefont {Scuseria}},
  \ and\ \bibinfo {author} {\bibfnamefont {M.}~\bibnamefont {Ernzerhof}},\
  }\href {\doibase 10.1063/1.1564060} {\bibfield  {journal} {\bibinfo
  {journal} {J. Chem. Phys.}\ }\textbf {\bibinfo {volume} {118}},\ \bibinfo
  {pages} {8207} (\bibinfo {year} {2003})}\BibitemShut {NoStop}%
\bibitem [{\citenamefont {Heyd}\ \emph {et~al.}(2006)\citenamefont {Heyd},
  \citenamefont {Scuseria},\ and\ \citenamefont {Ernzerhof}}]{HSE06}%
  \BibitemOpen
  \bibfield  {author} {\bibinfo {author} {\bibfnamefont {J.}~\bibnamefont
  {Heyd}}, \bibinfo {author} {\bibfnamefont {G.~E.}\ \bibnamefont {Scuseria}},
  \ and\ \bibinfo {author} {\bibfnamefont {M.}~\bibnamefont {Ernzerhof}},\
  }\href {\doibase 10.1063/1.2204597} {\bibfield  {journal} {\bibinfo
  {journal} {J. Chem. Phys.}\ }\textbf {\bibinfo {volume} {124}},\ \bibinfo
  {eid} {219906} (\bibinfo {year} {2006})}\BibitemShut {NoStop}%
\bibitem [{\citenamefont {Kresse}\ and\ \citenamefont
  {Furthm{\"u}ller}(1996)}]{Kressevasp1}%
  \BibitemOpen
  \bibfield  {author} {\bibinfo {author} {\bibfnamefont {G.}~\bibnamefont
  {Kresse}}\ and\ \bibinfo {author} {\bibfnamefont {J.}~\bibnamefont
  {Furthm{\"u}ller}},\ }\href {\doibase 10.1016/0927-0256(96)00008-0}
  {\bibfield  {journal} {\bibinfo  {journal} {Computational Materials Science}\
  }\textbf {\bibinfo {volume} {6}},\ \bibinfo {pages} {15} (\bibinfo {year}
  {1996})}\BibitemShut {NoStop}%
\bibitem [{\citenamefont {Kresse}\ and\ \citenamefont
  {Joubert}(1999)}]{Kressevasp3}%
  \BibitemOpen
  \bibfield  {author} {\bibinfo {author} {\bibfnamefont {G.}~\bibnamefont
  {Kresse}}\ and\ \bibinfo {author} {\bibfnamefont {D.}~\bibnamefont
  {Joubert}},\ }\href {\doibase 10.1103/PhysRevB.59.1758} {\bibfield  {journal}
  {\bibinfo  {journal} {Phys. Rev. B}\ }\textbf {\bibinfo {volume} {59}},\
  \bibinfo {pages} {1758} (\bibinfo {year} {1999})}\BibitemShut {NoStop}%
\bibitem [{vas()}]{vasp}%
  \BibitemOpen
  \href@noop {} {}\bibinfo {howpublished}
  {\url{https://www.vasp.at/}}\BibitemShut {NoStop}%
\bibitem [{\citenamefont {Pickard}\ and\ \citenamefont
  {Mauri}(2001)}]{Pickard01}%
  \BibitemOpen
  \bibfield  {author} {\bibinfo {author} {\bibfnamefont {C.~J.}\ \bibnamefont
  {Pickard}}\ and\ \bibinfo {author} {\bibfnamefont {F.}~\bibnamefont
  {Mauri}},\ }\href {\doibase 10.1103/PhysRevB.63.245101} {\bibfield  {journal}
  {\bibinfo  {journal} {Phys. Rev. B}\ }\textbf {\bibinfo {volume} {63}},\
  \bibinfo {pages} {245101} (\bibinfo {year} {2001})}\BibitemShut {NoStop}%
\bibitem [{\citenamefont {Pickard}\ and\ \citenamefont
  {Mauri}(2002)}]{Pickard02}%
  \BibitemOpen
  \bibfield  {author} {\bibinfo {author} {\bibfnamefont {C.~J.}\ \bibnamefont
  {Pickard}}\ and\ \bibinfo {author} {\bibfnamefont {F.}~\bibnamefont
  {Mauri}},\ }\href {\doibase 10.1103/PhysRevLett.88.086403} {\bibfield
  {journal} {\bibinfo  {journal} {Phys. Rev. Lett.}\ }\textbf {\bibinfo
  {volume} {88}},\ \bibinfo {pages} {086403} (\bibinfo {year}
  {2002})}\BibitemShut {NoStop}%
\bibitem [{\citenamefont {Giannozzi}\ \emph {et~al.}(2009)\citenamefont
  {Giannozzi}, \citenamefont {Baroni}, \citenamefont {Bonini}, \citenamefont
  {Calandra}, \citenamefont {Car}, \citenamefont {Cavazzoni}, \citenamefont
  {Ceresoli}, \citenamefont {Chiarotti}, \citenamefont {Cococcioni},
  \citenamefont {Dabo}, \citenamefont {{Dal Corso}}, \citenamefont
  {de~Gironcoli}, \citenamefont {Fabris}, \citenamefont {Fratesi},
  \citenamefont {Gebauer}, \citenamefont {Gerstmann}, \citenamefont
  {Gougoussis}, \citenamefont {Kokalj}, \citenamefont {Lazzeri}, \citenamefont
  {Martin-Samos}, \citenamefont {Marzari}, \citenamefont {Mauri}, \citenamefont
  {Mazzarello}, \citenamefont {Paolini}, \citenamefont {Pasquarello},
  \citenamefont {Paulatto}, \citenamefont {Sbraccia}, \citenamefont {Scandolo},
  \citenamefont {Sclauzero}, \citenamefont {Seitsonen}, \citenamefont
  {Smogunov}, \citenamefont {Umari},\ and\ \citenamefont
  {Wentzcovitch}}]{QE-2009}%
  \BibitemOpen
  \bibfield  {author} {\bibinfo {author} {\bibfnamefont {P.}~\bibnamefont
  {Giannozzi}}, \bibinfo {author} {\bibfnamefont {S.}~\bibnamefont {Baroni}},
  \bibinfo {author} {\bibfnamefont {N.}~\bibnamefont {Bonini}}, \bibinfo
  {author} {\bibfnamefont {M.}~\bibnamefont {Calandra}}, \bibinfo {author}
  {\bibfnamefont {R.}~\bibnamefont {Car}}, \bibinfo {author} {\bibfnamefont
  {C.}~\bibnamefont {Cavazzoni}}, \bibinfo {author} {\bibfnamefont
  {D.}~\bibnamefont {Ceresoli}}, \bibinfo {author} {\bibfnamefont {G.~L.}\
  \bibnamefont {Chiarotti}}, \bibinfo {author} {\bibfnamefont {M.}~\bibnamefont
  {Cococcioni}}, \bibinfo {author} {\bibfnamefont {I.}~\bibnamefont {Dabo}},
  \bibinfo {author} {\bibfnamefont {A.}~\bibnamefont {{Dal Corso}}}, \bibinfo
  {author} {\bibfnamefont {S.}~\bibnamefont {de~Gironcoli}}, \bibinfo {author}
  {\bibfnamefont {S.}~\bibnamefont {Fabris}}, \bibinfo {author} {\bibfnamefont
  {G.}~\bibnamefont {Fratesi}}, \bibinfo {author} {\bibfnamefont
  {R.}~\bibnamefont {Gebauer}}, \bibinfo {author} {\bibfnamefont
  {U.}~\bibnamefont {Gerstmann}}, \bibinfo {author} {\bibfnamefont
  {C.}~\bibnamefont {Gougoussis}}, \bibinfo {author} {\bibfnamefont
  {A.}~\bibnamefont {Kokalj}}, \bibinfo {author} {\bibfnamefont
  {M.}~\bibnamefont {Lazzeri}}, \bibinfo {author} {\bibfnamefont
  {L.}~\bibnamefont {Martin-Samos}}, \bibinfo {author} {\bibfnamefont
  {N.}~\bibnamefont {Marzari}}, \bibinfo {author} {\bibfnamefont
  {F.}~\bibnamefont {Mauri}}, \bibinfo {author} {\bibfnamefont
  {R.}~\bibnamefont {Mazzarello}}, \bibinfo {author} {\bibfnamefont
  {S.}~\bibnamefont {Paolini}}, \bibinfo {author} {\bibfnamefont
  {A.}~\bibnamefont {Pasquarello}}, \bibinfo {author} {\bibfnamefont
  {L.}~\bibnamefont {Paulatto}}, \bibinfo {author} {\bibfnamefont
  {C.}~\bibnamefont {Sbraccia}}, \bibinfo {author} {\bibfnamefont
  {S.}~\bibnamefont {Scandolo}}, \bibinfo {author} {\bibfnamefont
  {G.}~\bibnamefont {Sclauzero}}, \bibinfo {author} {\bibfnamefont {A.~P.}\
  \bibnamefont {Seitsonen}}, \bibinfo {author} {\bibfnamefont {A.}~\bibnamefont
  {Smogunov}}, \bibinfo {author} {\bibfnamefont {P.}~\bibnamefont {Umari}}, \
  and\ \bibinfo {author} {\bibfnamefont {R.~M.}\ \bibnamefont {Wentzcovitch}},\
  }\href {http://www.quantum-espresso.org} {\bibfield  {journal} {\bibinfo
  {journal} {Journal of Physics: Condensed Matter}\ }\textbf {\bibinfo {volume}
  {21}},\ \bibinfo {pages} {395502 (19pp)} (\bibinfo {year}
  {2009})}\BibitemShut {NoStop}%
\bibitem [{\citenamefont {Perdew}\ \emph
  {et~al.}(1996{\natexlab{b}})\citenamefont {Perdew}, \citenamefont {Burke},\
  and\ \citenamefont {Ernzerhof}}]{PBE}%
  \BibitemOpen
  \bibfield  {author} {\bibinfo {author} {\bibfnamefont {J.~P.}\ \bibnamefont
  {Perdew}}, \bibinfo {author} {\bibfnamefont {K.}~\bibnamefont {Burke}}, \
  and\ \bibinfo {author} {\bibfnamefont {M.}~\bibnamefont {Ernzerhof}},\ }\href
  {\doibase 10.1103/PhysRevLett.77.3865} {\bibfield  {journal} {\bibinfo
  {journal} {Phys. Rev. Lett.}\ }\textbf {\bibinfo {volume} {77}},\ \bibinfo
  {pages} {3865} (\bibinfo {year} {1996}{\natexlab{b}})}\BibitemShut {NoStop}%
\bibitem [{\citenamefont {von Bardeleben}\ \emph {et~al.}(2012)\citenamefont
  {von Bardeleben}, \citenamefont {Cantin}, \citenamefont {Gerstmann},
  \citenamefont {Scholle}, \citenamefont {Greulich-Weber}, \citenamefont
  {Rauls}, \citenamefont {Landmann}, \citenamefont {Schmidt}, \citenamefont
  {Gentils}, \citenamefont {Botsoa},\ and\ \citenamefont
  {Barthe}}]{Bardeleben12prl}%
  \BibitemOpen
  \bibfield  {author} {\bibinfo {author} {\bibfnamefont {H.~J.}\ \bibnamefont
  {von Bardeleben}}, \bibinfo {author} {\bibfnamefont {J.~L.}\ \bibnamefont
  {Cantin}}, \bibinfo {author} {\bibfnamefont {U.}~\bibnamefont {Gerstmann}},
  \bibinfo {author} {\bibfnamefont {A.}~\bibnamefont {Scholle}}, \bibinfo
  {author} {\bibfnamefont {S.}~\bibnamefont {Greulich-Weber}}, \bibinfo
  {author} {\bibfnamefont {E.}~\bibnamefont {Rauls}}, \bibinfo {author}
  {\bibfnamefont {M.}~\bibnamefont {Landmann}}, \bibinfo {author}
  {\bibfnamefont {W.~G.}\ \bibnamefont {Schmidt}}, \bibinfo {author}
  {\bibfnamefont {A.}~\bibnamefont {Gentils}}, \bibinfo {author} {\bibfnamefont
  {J.}~\bibnamefont {Botsoa}}, \ and\ \bibinfo {author} {\bibfnamefont {M.~F.}\
  \bibnamefont {Barthe}},\ }\href {\doibase 10.1103/PhysRevLett.109.206402}
  {\bibfield  {journal} {\bibinfo  {journal} {Phys. Rev. Lett.}\ }\textbf
  {\bibinfo {volume} {109}},\ \bibinfo {pages} {206402} (\bibinfo {year}
  {2012})}\BibitemShut {NoStop}%
\bibitem [{\citenamefont {von Bardeleben}\ \emph {et~al.}(2014)\citenamefont
  {von Bardeleben}, \citenamefont {Cantin}, \citenamefont {Vrielinck},
  \citenamefont {Callens}, \citenamefont {Binet}, \citenamefont {Rauls},\ and\
  \citenamefont {Gerstmann}}]{Bardeleben14}%
  \BibitemOpen
  \bibfield  {author} {\bibinfo {author} {\bibfnamefont {H.~J.}\ \bibnamefont
  {von Bardeleben}}, \bibinfo {author} {\bibfnamefont {J.~L.}\ \bibnamefont
  {Cantin}}, \bibinfo {author} {\bibfnamefont {H.}~\bibnamefont {Vrielinck}},
  \bibinfo {author} {\bibfnamefont {F.}~\bibnamefont {Callens}}, \bibinfo
  {author} {\bibfnamefont {L.}~\bibnamefont {Binet}}, \bibinfo {author}
  {\bibfnamefont {E.}~\bibnamefont {Rauls}}, \ and\ \bibinfo {author}
  {\bibfnamefont {U.}~\bibnamefont {Gerstmann}},\ }\href {\doibase
  10.1103/PhysRevB.90.085203} {\bibfield  {journal} {\bibinfo  {journal} {Phys.
  Rev. B}\ }\textbf {\bibinfo {volume} {90}},\ \bibinfo {pages} {085203}
  (\bibinfo {year} {2014})}\BibitemShut {NoStop}%
\bibitem [{\citenamefont {Pfanner}\ \emph {et~al.}(2012)\citenamefont
  {Pfanner}, \citenamefont {Freysoldt}, \citenamefont {Neugebauer},\ and\
  \citenamefont {Gerstmann}}]{Pfanner12}%
  \BibitemOpen
  \bibfield  {author} {\bibinfo {author} {\bibfnamefont {G.}~\bibnamefont
  {Pfanner}}, \bibinfo {author} {\bibfnamefont {C.}~\bibnamefont {Freysoldt}},
  \bibinfo {author} {\bibfnamefont {J.}~\bibnamefont {Neugebauer}}, \ and\
  \bibinfo {author} {\bibfnamefont {U.}~\bibnamefont {Gerstmann}},\ }\href
  {\doibase 10.1103/PhysRevB.85.195202} {\bibfield  {journal} {\bibinfo
  {journal} {Phys. Rev. B}\ }\textbf {\bibinfo {volume} {85}},\ \bibinfo
  {pages} {195202} (\bibinfo {year} {2012})}\BibitemShut {NoStop}%
\bibitem [{\citenamefont {George}\ \emph {et~al.}(2013)\citenamefont {George},
  \citenamefont {Behrends}, \citenamefont {Schnegg}, \citenamefont {Schulze},
  \citenamefont {Fehr}, \citenamefont {Korte}, \citenamefont {Rech},
  \citenamefont {Lips}, \citenamefont {Rohrm{\"u}ller}, \citenamefont {Rauls},
  \citenamefont {Schmidt},\ and\ \citenamefont {Gerstmann}}]{George13}%
  \BibitemOpen
  \bibfield  {author} {\bibinfo {author} {\bibfnamefont {B.~M.}\ \bibnamefont
  {George}}, \bibinfo {author} {\bibfnamefont {J.}~\bibnamefont {Behrends}},
  \bibinfo {author} {\bibfnamefont {A.}~\bibnamefont {Schnegg}}, \bibinfo
  {author} {\bibfnamefont {T.~F.}\ \bibnamefont {Schulze}}, \bibinfo {author}
  {\bibfnamefont {M.}~\bibnamefont {Fehr}}, \bibinfo {author} {\bibfnamefont
  {L.}~\bibnamefont {Korte}}, \bibinfo {author} {\bibfnamefont
  {B.}~\bibnamefont {Rech}}, \bibinfo {author} {\bibfnamefont {K.}~\bibnamefont
  {Lips}}, \bibinfo {author} {\bibfnamefont {M.}~\bibnamefont
  {Rohrm{\"u}ller}}, \bibinfo {author} {\bibfnamefont {E.}~\bibnamefont
  {Rauls}}, \bibinfo {author} {\bibfnamefont {W.~G.}\ \bibnamefont {Schmidt}},
  \ and\ \bibinfo {author} {\bibfnamefont {U.}~\bibnamefont {Gerstmann}},\
  }\href {\doibase 10.1103/PhysRevLett.110.136803} {\bibfield  {journal}
  {\bibinfo  {journal} {Phys. Rev. Lett.}\ }\textbf {\bibinfo {volume} {110}},\
  \bibinfo {pages} {136803} (\bibinfo {year} {2013})}\BibitemShut {NoStop}%
\bibitem [{\citenamefont {Rohrm{\"u}ller}\ \emph {et~al.}(2017)\citenamefont
  {Rohrm{\"u}ller}, \citenamefont {Schmidt},\ and\ \citenamefont
  {Gerstmann}}]{Rohrmuller17}%
  \BibitemOpen
  \bibfield  {author} {\bibinfo {author} {\bibfnamefont {M.}~\bibnamefont
  {Rohrm{\"u}ller}}, \bibinfo {author} {\bibfnamefont {W.~G.}\ \bibnamefont
  {Schmidt}}, \ and\ \bibinfo {author} {\bibfnamefont {U.}~\bibnamefont
  {Gerstmann}},\ }\href {\doibase 10.1103/PhysRevB.95.125310} {\bibfield
  {journal} {\bibinfo  {journal} {Phys. Rev. B}\ }\textbf {\bibinfo {volume}
  {95}},\ \bibinfo {pages} {125310} (\bibinfo {year} {2017})}\BibitemShut
  {NoStop}%
\bibitem [{\citenamefont {Cococcioni}\ and\ \citenamefont
  {de~Gironcoli}(2005)}]{Cococcioni5}%
  \BibitemOpen
  \bibfield  {author} {\bibinfo {author} {\bibfnamefont {M.}~\bibnamefont
  {Cococcioni}}\ and\ \bibinfo {author} {\bibfnamefont {S.}~\bibnamefont
  {de~Gironcoli}},\ }\href {\doibase 10.1103/PhysRevB.71.035105} {\bibfield
  {journal} {\bibinfo  {journal} {Phys. Rev. B}\ }\textbf {\bibinfo {volume}
  {71}},\ \bibinfo {pages} {035105} (\bibinfo {year} {2005})}\BibitemShut
  {NoStop}%
\bibitem [{\citenamefont {Timrov}\ \emph {et~al.}(2018)\citenamefont {Timrov},
  \citenamefont {Marzari},\ and\ \citenamefont {Cococcioni}}]{Timrov18}%
  \BibitemOpen
  \bibfield  {author} {\bibinfo {author} {\bibfnamefont {I.}~\bibnamefont
  {Timrov}}, \bibinfo {author} {\bibfnamefont {N.}~\bibnamefont {Marzari}}, \
  and\ \bibinfo {author} {\bibfnamefont {M.}~\bibnamefont {Cococcioni}},\
  }\href {\doibase 10.1103/PhysRevB.98.085127} {\bibfield  {journal} {\bibinfo
  {journal} {Phys. Rev. B}\ }\textbf {\bibinfo {volume} {98}},\ \bibinfo
  {pages} {085127} (\bibinfo {year} {2018})}\BibitemShut {NoStop}%
\bibitem [{\citenamefont {Ho}\ \emph {et~al.}(2018)\citenamefont {Ho},
  \citenamefont {Frauenheim},\ and\ \citenamefont {De\'ak}}]{Ho18}%
  \BibitemOpen
  \bibfield  {author} {\bibinfo {author} {\bibfnamefont {Q.~D.}\ \bibnamefont
  {Ho}}, \bibinfo {author} {\bibfnamefont {T.}~\bibnamefont {Frauenheim}}, \
  and\ \bibinfo {author} {\bibfnamefont {P.}~\bibnamefont {De\'ak}},\ }\href
  {\doibase 10.1063/1.5049861} {\bibfield  {journal} {\bibinfo  {journal}
  {Journal of Applied Physics}\ }\textbf {\bibinfo {volume} {124}},\ \bibinfo
  {pages} {145702} (\bibinfo {year} {2018})},\ \Eprint
  {http://arxiv.org/abs/https://doi.org/10.1063/1.5049861}
  {https://doi.org/10.1063/1.5049861} \BibitemShut {NoStop}%
\bibitem [{\citenamefont {Ham}(1965)}]{Ham65}%
  \BibitemOpen
  \bibfield  {author} {\bibinfo {author} {\bibfnamefont {F.~S.}\ \bibnamefont
  {Ham}},\ }\href {\doibase 10.1103/PhysRev.138.A1727} {\bibfield  {journal}
  {\bibinfo  {journal} {Phys. Rev.}\ }\textbf {\bibinfo {volume} {138}},\
  \bibinfo {pages} {A1727} (\bibinfo {year} {1965})}\BibitemShut {NoStop}%
\bibitem [{\citenamefont {Mauger}\ \emph {et~al.}(1987)\citenamefont {Mauger},
  \citenamefont {von Bardeleben}, \citenamefont {Bourgoin},\ and\ \citenamefont
  {Lannoo}}]{Mauger87}%
  \BibitemOpen
  \bibfield  {author} {\bibinfo {author} {\bibfnamefont {A.}~\bibnamefont
  {Mauger}}, \bibinfo {author} {\bibfnamefont {H.~J.}\ \bibnamefont {von
  Bardeleben}}, \bibinfo {author} {\bibfnamefont {J.~C.}\ \bibnamefont
  {Bourgoin}}, \ and\ \bibinfo {author} {\bibfnamefont {M.}~\bibnamefont
  {Lannoo}},\ }\href {\doibase 10.1103/PhysRevB.36.5982} {\bibfield  {journal}
  {\bibinfo  {journal} {Phys. Rev. B}\ }\textbf {\bibinfo {volume} {36}},\
  \bibinfo {pages} {5982} (\bibinfo {year} {1987})}\BibitemShut {NoStop}%
\bibitem [{gip()}]{gipaw}%
  \BibitemOpen
  \href@noop {} {}\bibinfo {howpublished}
  {\url{http://qe-forge.org/gf/project/qe-gipaw/}}\BibitemShut {NoStop}%
\bibitem [{\citenamefont {Bl\"ugel}\ \emph {et~al.}(1987)\citenamefont
  {Bl\"ugel}, \citenamefont {Akai}, \citenamefont {Zeller},\ and\ \citenamefont
  {Dederichs}}]{Blugel87}%
  \BibitemOpen
  \bibfield  {author} {\bibinfo {author} {\bibfnamefont {S.}~\bibnamefont
  {Bl\"ugel}}, \bibinfo {author} {\bibfnamefont {H.}~\bibnamefont {Akai}},
  \bibinfo {author} {\bibfnamefont {R.}~\bibnamefont {Zeller}}, \ and\ \bibinfo
  {author} {\bibfnamefont {P.~H.}\ \bibnamefont {Dederichs}},\ }\href {\doibase
  10.1103/PhysRevB.35.3271} {\bibfield  {journal} {\bibinfo  {journal} {Phys.
  Rev. B}\ }\textbf {\bibinfo {volume} {35}},\ \bibinfo {pages} {3271}
  (\bibinfo {year} {1987})}\BibitemShut {NoStop}%
\bibitem [{\citenamefont {Van~de Walle}\ and\ \citenamefont
  {Bl\"ochl}(1993)}]{VdWalleBlochl93}%
  \BibitemOpen
  \bibfield  {author} {\bibinfo {author} {\bibfnamefont {C.~G.}\ \bibnamefont
  {Van~de Walle}}\ and\ \bibinfo {author} {\bibfnamefont {P.~E.}\ \bibnamefont
  {Bl\"ochl}},\ }\href {\doibase 10.1103/PhysRevB.47.4244} {\bibfield
  {journal} {\bibinfo  {journal} {Phys. Rev. B}\ }\textbf {\bibinfo {volume}
  {47}},\ \bibinfo {pages} {4244} (\bibinfo {year} {1993})}\BibitemShut
  {NoStop}%
\bibitem [{\citenamefont {Kyrtsos}\ \emph {et~al.}(2017)\citenamefont
  {Kyrtsos}, \citenamefont {Matsubara},\ and\ \citenamefont
  {Bellotti}}]{Kyrtsos17}%
  \BibitemOpen
  \bibfield  {author} {\bibinfo {author} {\bibfnamefont {A.}~\bibnamefont
  {Kyrtsos}}, \bibinfo {author} {\bibfnamefont {M.}~\bibnamefont {Matsubara}},
  \ and\ \bibinfo {author} {\bibfnamefont {E.}~\bibnamefont {Bellotti}},\
  }\href {\doibase 10.1103/PhysRevB.95.245202} {\bibfield  {journal} {\bibinfo
  {journal} {Phys. Rev. B}\ }\textbf {\bibinfo {volume} {95}},\ \bibinfo
  {pages} {245202} (\bibinfo {year} {2017})}\BibitemShut {NoStop}%
\bibitem [{sup()}]{supinfo}%
  \BibitemOpen
  \href@noop {} {}\bibinfo {note} {The Supplementary Information contains
  additional figures of the spin density, $g$-tensor and large hyperfine atoms
  for some of the models, not shown in the main text. Also VESTA files which
  allow to view these structures from different angles are
  provided.}\BibitemShut {Stop}%
\bibitem [{\citenamefont {Riplinger}\ \emph {et~al.}(2009)\citenamefont
  {Riplinger}, \citenamefont {Kao}, \citenamefont {Rosen}, \citenamefont
  {Kathirvelu}, \citenamefont {Eaton}, \citenamefont {Eaton}, \citenamefont
  {Kutateladze},\ and\ \citenamefont {Neese}}]{Riplinger}%
  \BibitemOpen
  \bibfield  {author} {\bibinfo {author} {\bibfnamefont {C.}~\bibnamefont
  {Riplinger}}, \bibinfo {author} {\bibfnamefont {J.~P.~Y.}\ \bibnamefont
  {Kao}}, \bibinfo {author} {\bibfnamefont {G.~M.}\ \bibnamefont {Rosen}},
  \bibinfo {author} {\bibfnamefont {V.}~\bibnamefont {Kathirvelu}}, \bibinfo
  {author} {\bibfnamefont {G.~R.}\ \bibnamefont {Eaton}}, \bibinfo {author}
  {\bibfnamefont {S.~S.}\ \bibnamefont {Eaton}}, \bibinfo {author}
  {\bibfnamefont {A.}~\bibnamefont {Kutateladze}}, \ and\ \bibinfo {author}
  {\bibfnamefont {F.}~\bibnamefont {Neese}},\ }\href {\doibase
  10.1021/ja901150j} {\bibfield  {journal} {\bibinfo  {journal} {Journal of the
  American Chemical Society}\ }\textbf {\bibinfo {volume} {131}},\ \bibinfo
  {pages} {10092} (\bibinfo {year} {2009})},\ \bibinfo {note} {pMID:
  19621964},\ \Eprint {http://arxiv.org/abs/https://doi.org/10.1021/ja901150j}
  {https://doi.org/10.1021/ja901150j} \BibitemShut {NoStop}%
\end{thebibliography}%

\end{document}